\documentclass[11pt]{article}
\usepackage[margin=2cm]{geometry}
\usepackage{amsmath,amsthm,amssymb,amsfonts}
\usepackage[retainorgcmds]{IEEEtrantools}

\usepackage[]{natbib} 
\usepackage[pagebackref=false,
            linktoc=page,
            colorlinks=true,
            linkcolor=blue,
            citecolor=blue]{hyperref}
\usepackage{authblk}
\usepackage{pstool,psfrag}
\usepackage{bbm}
\usepackage{bm}
\usepackage[outdir=./]{epstopdf}
\usepackage{psfrag}
\usepackage{subcaption}
\usepackage{enumerate}
\usepackage{booktabs}
\usepackage[linesnumbered,lined,boxed,commentsnumbered]{algorithm2e} 
\usepackage{caption}
\usepackage[utf8]{inputenc}
\numberwithin{equation} {section}
\usepackage{graphicx}
\usepackage{mathtools}

\usepackage{geometry}
\usepackage[official]{eurosym} 
\usepackage{listings}
\mathchardef\mhyphen="2D  
\usepackage{color}

\numberwithin{Theorem}{section}
\numberwithin{Definition}{section}
\numberwithin{Lemma}{section}
\numberwithin{Algorithm}{section}
\numberwithin{equation}{section}

\theoremstyle{plain}
\newtheorem{theorem}{Theorem}[section]

\theoremstyle{definition}

\theoremstyle{remark}

\author[$\dagger$\footnote{Corresponding author}]{\textbf{Graeme Auld} }
\author[$\dagger\dagger$]{\textbf{Ioannis Papastathopoulos}}
\affil[$\dagger$]{\small Department of Mathematics and Computer Science,  Chulalongkorn University, Bangkok, Thailand}
\affil[$\dagger\dagger$]{\small School of Mathematics and Maxwell Institute,
  University of Edinburgh, Edinburgh, United Kingdom}
\affil[$$]{\small
  GraemeRoss.A@chula.ac.th 
  $\quad$ i.papastathopoulos@ed.ac.uk} 
\date{}
\begin{document}
\title{\textbf{Time series conditional extremes}}  \label{TSCE}

\maketitle

\begin{abstract} 
  \noindent Accurate modelling of the joint extremal dependence
  structure within a stationary time series is a challenging problem
  that is important in many applications.\ Several previous approaches
  to this problem are only applicable to certain types of extremal
  dependence in the time series such as asymptotic dependence, or to
  Markov time series of finite order.\ In this paper, we develop
  statistical methodology for time series extremes based on recent
  probabilistic results that allow us to flexibly model the decay of a
  stationary time series after witnessing an extreme event.\ While
  Markov sequences of finite order are naturally accommodated by our
  approach, we consider a broader setup, based on the conditional
  extreme value model, which allows for a wide range of possible
  dependence structures in the time series.\ We consider inference
  based on Monte Carlo simulation and derive an upper bound for the
  variance of a commonly used importance sampler.\ Our methodology is
  illustrated via estimation of cluster functionals in simulated data
  and in a time series of daily maximum temperatures from Orleans,
  France.
\end{abstract} {\bf Keywords:} conditional extremes, asymptotic independence, time series extremes, Markov chains
\section{Introduction} \label{Intro}
\label{sec:intro}

Many types of extreme events in nature derive their impact from the
occurrence of a cluster of extreme values, i.e., several extremely
large or small values are observed within a short period of time.\ A
flood may be the result of several days of heavy rainfall and a
heatwave the result of several days of high temperature.\ Such events
may lead to damage of important infrastructure, give rise to large
insurance claims and to a loss of human life.\ For example, as a
result of the devastating European heatwave in the summer of 2003, an
estimated 40,000-70,000 heat-related deaths were recorded
\citep{fischer10, robine08} with associated economic losses in excess
of \euro{13} billion \citep{bono04}.\ The development of statistical
models that can accurately replicate the extremal clustering behaviour
of a natural process requires us to understand the extremal dependence
structure of that process.

One way to understand how extremal dependence varies with lag in a
time series is via the tail dependence function $\chi$
\citep{ledtawn03}.\ If $\{X_n\}_{n\in\mathbb{Z}}$ is a stationary time
series with marginal distribution function $F$, then $\chi$ is defined
by
\begin{equation}
\chi_i = \lim_{x\to 1}\mathbb{P}(F(X_i)  > x \mid F(X_0) > x), \quad i\in \mathbb{N},
\end{equation}
provided the limit exists.\ When $\chi_i = 0$, we say that $X_0$ and
$X_i$ are asymptotically independent (AI) \citep{Sibuya60} in which
case $X_0$ and $X_i$ cannot take their largest values simultaneously.\
If $\chi_i = 0$ for all $i \geq 1,$ then we say that
$\{X_n\}_{n\in\mathbb{Z}}$ is an AI time series.\ Although the
extremes of an AI time series occur singly in the limit, strong
dependence may still exist at moderately extreme levels, a fact that
is not captured by the asymptotic measure $\chi$.\ When $\chi_i > 0$
for some $i \geq 1$, we say the time series is asymptotically
dependent.\ Several previous models for time series extremes, such as
those considered in \cite{Smith97}, \cite{perfekt97} and
\cite{basraksegers}, are appropriate only in the case of asymptotic
dependence.\ As illustrated in \cite{winttawn16} in the context of
modelling heatwaves, if an asymptotically dependent process is
incorrectly used for to model a process that exhibits asymptotic
independence, then the probability of events related to subsets of
variables attaining large values simultaneously may be severely
overestimated.\ As environmental time series often exhibit asymptotic
independence, it is important when modelling such processes to specify
models that are able to accommodate this extremal dependence class.
 
A powerful modelling framework that may be used to model all types of
extremal dependence is the conditional extremes model of
\cite{hefftawn04} which we discuss now only in the bivariate case for
simplicity.\ If $X_0$ and $X_1$ are exponential tailed random
variables, \cite{hefftawn04} show that for a broad class of dependence
structures on $(X_0, X_1)$, there exist location and scale norming
functions $a_1:\mathbb{R}\to\mathbb{R}$ and
$b_1:\mathbb{R}\to\mathbb{R}_+$ such that
\begin{equation}
\left. \bigg(X_0-u, \frac{X_1 - a_1(X_0)}{b_1(X_0)}\bigg)  \right\vert X_0 > u \overset{D}{\to} (E_0, Z^0),   \quad u\to\infty,  \label{HT_bivariate}
\end{equation}
where $E_0$ is a unit exponential random variable independent of
$Z^0$, and $Z^0 \sim G^0$, where $G^0$ is a non-degenerate
distribution function.\ Moreover, the norming functions $a_1$ and
$b_1$ may be identified as belonging to simple parametric families.\
In particular, when $X_0$ and $X_1$ are unit Laplace random variables,
then $a_1(x) = \alpha\,x$ and $b_1(x) = x^{\beta}$, where
$\alpha\in\,[-1,1]$ and $\beta\in\,(-\infty, 1)$
\citep{keefpaptawn13}.\ The cases $\alpha < 1$ and $\alpha = 1$
correspond to $X_0$ and $X_1$ being asymptotically independent and
asymptotically dependent, respectively.

\cite{winttawn16} use the conditional extremes approach to model
extremes in a time series of daily temperatures and estimate
probabilities of extreme events related to heatwaves.\ By making a
first-order Markov assumption and assuming that the dependence
structure of $(X_t, X_{t+1})$ belongs to the class identified by
\cite{hefftawn04}, they motivate a model for $X_{t+1}$ conditional on
$X_t > u$ for a large threshold $u$ as
\begin{equation}
X_{t+1} = \alpha\,X_t + X_t^{\beta}Z_{t+1} , \quad X_t>u, \label{WT_rec}
\end{equation}
with parameters $\alpha$ and $\beta$ to estimate.\ The random variable
$Z_{t+1}$ in model (\ref{WT_rec}) corresponds to the limiting random
variable $Z^0$ in the convergence (\ref{HT_bivariate}) whose
distribution is estimated non-parametrically.\ By simulating an
initial exceedance, $X_t$, of the threshold $u$ using the exponential
tailed assumption and iterativing forwards in time from
model(\ref{WT_rec}), they are able to simulate the behaviour of the
daily temperature series after entering an extreme state.\ When
applying the recurrence (\ref{WT_rec}), the residual $Z_{t+1}$ is
simulated independently of the initial exceedance $X_t$, and $Z_{t+i}$
is simulated independently of $Z_{t+j}$ for $i\neq
j$. \cite{Winter2017} generalise this model to higher-order Markov
time series.

In this paper we also consider time series models based on the
conditional extremes approach of \cite{hefftawn04} that allow us to
simulate the behaviour of both asymptotically independent and
asymptotically dependent stationary time series after a large
threshold is exceeded.\% Our framework is broader than that of
We consider modelling a large block of consecutive
observations after observing an exceedance of a large threshold, that
is, we consider models for $(X_{t+1}, X_{t+2},\ldots, X_{t+k})$,
conditional on $X_t>u$ for some large threshold $u,$ where $k$ is a
positive integer that may be reasonably large.\ The precise choice of
the constant $k$ will be context dependent and in practice will be
chosen to be sufficiently large so that the $k+1$ observations
$(X_t, X_{t+1},\ldots,X_{t+k})$ encompass the full duration of an
extreme event.\ For example, in the data application considered in
Section \ref{DataApp}, where we consider the behaviour of a daily
temperature time series over a three week period conditioned on a
large temperature at the start of the period, we select $k=20$.\ Our
motivation for this approach is that by fitting models whose input
data consists of a large block of observations encompassing the full
duration of an extreme event, we hope to be able to accurately model
the subsequent behaviour of a stationary time series after entering an
extreme state for a broad class of possible dependence structures.

We will assume that the copula of $(X_t, X_{t+1},\ldots,X_{t+k})$
belongs to the class identified by \cite{hefftawn04}.\ This motivates
a model for the conditional distribution of $X_{t+i}$ given $X_t > u$,
for an appropriately chosen large threshold $u$ as
\begin{equation}
  X_{t+i} = \alpha_i\,X_t + X_t^{\beta_i}Z_{t+i}, \quad 1 \leq i \leq k.  \label{Our_rec}
\end{equation} 
This modelling framework is also considered in \cite{eastoetawn12} and
is similar to that of \cite{Winter2017} where $k$ is taken to be the
order of the assumed Markov process.\ The novelty in our approach is
to consider models that impose structure on
$(\alpha_1, \alpha_2, \ldots,\alpha_k)$ and
$(\beta_1,\beta_2,\ldots,\beta_k),$ whereas in the models of
\cite{eastoetawn12} and \cite{Winter2017}, $\alpha_i$ and $\beta_i$
are completely uncoupled from $\alpha_j$ and $\beta_j$ when
$i\neq j$.\ A further novelty of our approach is that we may also
consider fully parametric models for the residual vector
$\bm{Z} = (Z_{t+1},\ldots,Z_{t+k})$ in the statistical model
(\ref{Our_rec}).

Our methodology is influenced by the approach of \cite{wadstawn22}
where, in a spatial setting, the authors consider simulating over a
spatial field conditional on a large observation at a reference
location.\ We build on their innovation of imposing structure on the
shape parameter of the residual process by also imposing structure in
the location and scale of the residuals.\ One advantage of working in
the simpler time series setting is that for a large class of
stationary Markov time series there are recent key probabilistic
results that allow us to impose theoretically justified structure on
model parameters \citep{papaetal17, papastawn20}.\ Although the
examples we consider focus on asymptotically independent Markov time
series, the methodology can be adapted to more general processes that
are neither Markov, nor asymptotically independent.

The structure of the paper is as follows.\ Section
\ref{NotAss_CondSec} introduces the notation and main assumptions that
are used throughout the rest of the paper.\ Section \ref{StatInf}
considers statistical modelling and approaches to statistical
inference.\ Section \ref{SimulationSec} presents simulation methods
that may be used together with the models from Section \ref{StatInf}
to estimate probabilities of extreme events, and we also prove the
consistency of the importance sampling algorithm of
\cite{wadstawn22}.\ Section \ref{Ex} considers examples of AI Markov
time series and assesses the behaviour of the models from Section
\ref{StatInf} in a simulation study.\ Section \ref{DataApp}
illustrates our methodology using a time series of daily maximum
temperature from Orleans, France.

\section{Notation and assumptions}   \label{NotAss_CondSec}

If $i$ and $j$ are two integers with $i \leq j$ we define the set
$i{\,:\,}j$ to be all integers from $i$ to $j$ inclusive, i.e.,
$i{\,:\,}j = \{n\in \mathbb{Z}: i \leq n \leq j \}$.\ Also, the block
of consecutive random variables $\{X_n: i \leq n \leq j \}$ is denoted
by $\bm{X}_{i\,:\,j}$.\ More generally, if $A$ is a set of positive
integers, the collection of random variables $\{X_i : i\in A\}$ is
denoted by $\bm{X}_A$.\ In what follows all arithmetic operations are
vectorized and such operations involving two vectors should be
interpreted componentwise.\ Also, operations involving vectors of
different lengths are defined by recycling the smaller vector to match
the length of the larger vector.\ So, for example, if $\alpha \geq 0$
and $k\in \mathbb{N}$ are constants and
$\bm{y} = (y_1,\ldots,y_d) \in \mathbb{R}^d$, then
$\alpha^{1\,:\,k} = (\alpha, \alpha^2,\ldots,\alpha^k)$ and
$\alpha + \bm{y} = (\alpha+y_1,\ldots,\alpha+y_d)$.

We will use the symbol $\sim$ to mean ``is distributed as'' where the
symbol to the right of $\sim$ may be either a distribution or density
function.\ If $\bm{X}$ is a random vector and $g$ a measurable
function, an expressions such as $\mathbb{E}_{\pi}\{g(\bm{X})\}$
denotes the expected value of $g(\bm{X})$ when $\bm{X} \sim \pi$.

Unless otherwise stated, $\{X_n \}_{n\in\mathbb{Z}}$ will denote a
stationary time series with Laplace marginal distributions so that
\begin{equation} \label{LapMarg}
\mathbb{P}(X_n \leq x) =
\begin{cases}
\exp(x)/2, \quad x \leq 0, \\
1 - \exp(-x)/2, \quad x > 0
\end{cases}
\end{equation}
for each $n\in\mathbb{Z}$ and $x \in \mathbb{R}$.\ There is no loss in
generality in assuming a prescribed marginal distribution, since, any
stationary time series with continuous margins may be transformed to
another stationary time series with different continuous margins via
the probability integral transform.

We define the random set of times of exceedance of the threshold
$u > 0$ to be the set
\begin{equation}
T_u = \{ i\in \mathbb{N} : X_i > u\}.
\end{equation}
Our main assumption concerns the limiting behaviour of the time series
$\{X_n \}_{n\in\mathbb{Z}}$ conditional upon observing an extreme
state at an arbitrary position in the series, $X_0$ say.\ In
particular, we assume that for each finite set
$A \subset \mathbb{Z}\setminus\,\{0\}$, there exist location and scale
functions $\bm{a}_A =\{a_i: \mathbb{R} \to \mathbb{R}\}_{i\in\,A}$ and
$\bm{b}_A = \{b_i: \mathbb{R} \to \mathbb{R_+}\}_{i\in\mathbb{A}}$
respectively, such that
\begin{equation}
\left. \bigg(X_0 - u, \frac{\bm{X}_{A} - \bm{a}_{A}(X_0)}{\bm{b}_{A}(X_0)}\bigg) \right\vert X_0 > u \xrightarrow[]{D} (E_0, \bm{Z}_A)  \label{ForwardAssumption}
\end{equation}
as $u\to\infty,$ where $E_0$ is a unit exponential random variable
independent of the random vector $\bm{Z}_{A} \sim G_{A},$ where
$G_{A}$ is a joint distribution function on $\mathbb{R}^{|A|}$ with
non-degenerate margins that place no mass at $+\infty$.\ The
assumption (\ref{ForwardAssumption}) simply states that the finite
dimensional copulas of the process $\{X_n \}_{n\in\mathbb{Z}}$ belong
to the class identified by \cite{hefftawn04}.\ The Heffernan--Tawn
class is broad and includes all of the copula examples considered in
\cite{joe97}.\ 

Throughout most of the paper we will consider the simple special case
where $A = 1{\,:\,}k$ for some $k\in\mathbb{N},$ i.e., we consider the
behaviour of the process immediately after entering an extreme state.\
We will sometimes refer to this approach as the $k$-steps method as it
will allow us to build models for the time series up to $k$ time steps
forward from an initial threshold exceedance.\ The constant $k$ will
be chosen by the statistician depending on context and the sensitivity
of model parameter estimates to the choice of $k$ is considered in
Section \ref{Ex}.\ We believe this simple special case will be of most
frequent interest, although in Section \ref{ImpSamp} we describe
scenarios in which we need to be able to simulate both forwards and
backwards in time from an extreme event so that more general sets $A$
are required.

\section{Statistical modelling}  \label{StatInf}

\subsection{Marginal model}   \label{MargMod}

Although in equation (\ref{LapMarg}) we assume standard Laplace
marginal distributions, data encountered in applications will
typically not be on this scale and so a marginal transformation is
required to apply the models that follow.\ If
$\{Y_n\}_{n\in\mathbb{Z}}$ is a stationary time series with marginal
distribution function $F$, then the time series
$\{X_n\}_{n\in\mathbb{Z}}$ defined by
\begin{equation}
X_n = \begin{cases} 
\text{log}\,\{2F(Y_n)\}, \quad & \text{if\,\,} F(Y_n) < 1/2, \\
-\text{log}\,[2\{1 - F(Y_n)\}] ,  \quad & \text{if\,\,} F(Y_n) \geq 1/2,  \label{MarginTrans}
\end{cases}
\end{equation}
is stationary and has Laplace marginal distributions.\ In practice,
the exact form of $F$ will not be known and must be estimated.\ We use
a standard semi-parametric approach \citep{coletawn91} to estimate $F$
by the empirical distribution function below some high threshold
$u^{*}$ and via a generalized Pareto distribution (GPD) above
$u^{*}$.\ Thus we assume a model for the upper tail of $Y$ of the form
\begin{equation}
\mathbb{P}(Y - u^{*} > y \mid Y > u^{*}) = (1 + \xi\,y/\sigma)^{-1/\xi}_+,  \quad y > 0, \label{GPD}
\end{equation}
where $y_+ = \max\{y, 0\}$.\ The parameters $\sigma > 0$ and $\xi$
control the scale and shape of the tail of $Y$ respectively.\ We
estimate $\sigma$ and $\xi$ in equation (\ref{GPD}) by maximum
likelihood using the excesses $Y_t - u^{*}$ for each $t$ such that
$Y_t > u^{*}$ \citep{davismit90}.\ The threshold $u^*$ is selected by
fitting the GPD model for a range of high thresholds and selecting the
smallest threshold such that the estimates of $\sigma$ and $\xi$
stabilise.

Given the maximum likelihood estimators $\hat{\sigma}$ and $\hat{\xi}$, of $\sigma $ and $\xi$, an estimator of the marginal distribution $F$ based on a sample $\bm{Y}_{1\,:\,N}$ is then
\begin{equation}  \label{MargModSemiParam}
\hat{F}(y) = \begin{cases}
 N^{-1}\sum_{i=1}^N \mathbbm{1}[Y_i \leq y],  \quad & y \leq u^*,   \\
1 - \hat{p}_{u^{*}}\{1 + \hat{\xi}\,(y-y^{*}\}/\hat{\sigma})^{-1/\hat{\xi}}_+,  \quad &  y > u^*,
\end{cases}
\end{equation}
where $\hat{p}_{u^{*}} = N^{-1}\sum_{i=1}^{N}\mathbbm{1}[Y_i > u^*]$
is the empirical estimator of $Y_i$ exceeding the threshold $u^*$.

\subsection{Parametric models for norming functions}
We will use assumption (\ref{ForwardAssumption}) as our basis for
modelling the conditional distribution of $\bm{X}_{t+1\,:\,t+k}$ given
$X_t > u$.\ We will assume that the positive integer $k$ which
determines how many lags forward from the extreme event we wish to
model has been fixed and discuss this issue further in Section
\ref{Ex}.\ We will also assume that the threshold $u$ has been chosen
to be suitably large so that the convergence (\ref{ForwardAssumption})
holds as an equality exactly above $u$.\ By stationarity, the same
functions $\bm{a}_{1\,:\,k}$ and $\bm{b}_{1\,:\,k}$ can be used to
normalize $\bm{X}_{t+1\,:\,t+k}$ given $X_t > u$, as for
$\bm{X}_{1\,:\,k}$ given $X_0 > u$.\ Hence, we assume that
\begin{equation}
  \frac{\bm{X}_{t+1\,:\,t+k} - \bm{a}_{1\,:\,k}(X_t)}{\bm{b}_{1\,:\,k}(X_t)}= (Z_{t+1\mid t},\ldots,Z_{t+k\mid t}) =  \bm Z_{t+1\,:\,t+k\mid t}, \quad t \in T_u, \label{Norming}
\end{equation}
where the margins of $\bm Z_{t+1\,:\,t+k\mid t}$ are non-degenerate.\
As
$u \to \infty$, $\bm Z_{t+1\,:\,t+k\mid t}$ corresponds to a block of
length $k$ from the hidden tail chain, however as in our applications
$u$ will be finite, we will simply refer to
$\bm Z_{t+1\,:\,t+k\mid t}$ as a residual vector.\ We will consider
two approaches to statistical inference which differ in the manner in
which the residual vector $\bm Z_{t+1\,:\,t+k\mid t}$ is modelled.\ In
the first case, discussed in Section \ref{SemiParamSec},
$\bm Z_{t+1\,:\,t+k\mid t}$ is modelled non-parametrically, whereas in
Section \ref{ParamSec}, parametric models are discussed.

In order to make statistical inference tractable, we are required to
specify the forms of the norming functions $\bm{a}_{1\,:\,k}$ and
$\bm{b}_{1\,:\,k}$ and we consider two possibilities.\ Our first model
is motivated by the normings found in \cite{hefftawn04} which we
specify as
\begin{align}
  & \textbf{Model 1:}\quad a_i(x) = \alpha_ix, \quad b_i(x) = x^{\beta}, \quad i\in 1{\,:\,}k,\,\, \alpha_i \in [-1, 1],\,\,  \beta \in [0, 1).  \label{Mod1Norm}
    \intertext{
    A more flexible model would be to specify $b_i(x) = x^{\beta_i}$, i.e., to have different scale normalizations at each lag. We opt for the simple case
    where $\beta_i = \beta$ for each $i\in\,1{\,:\,}k$, in part due to a lack of useful theoretical results that can be used to impose structure on $\beta_i$
    and in part due to results in \cite{papaetal17} and \cite{papastawn20} that show that for Markov time series we may take $\beta_i = \beta$.\
    Our parameter restrictions on $\beta$ are stricter than those in \cite{hefftawn04}, where $\beta \in (-\infty, 1)$.\ 
    For negative values of $\beta,$ the relationship between $X_t$ and $X_{t+i}$, conditional on $X_t > u$, becomes deterministic as $u\to \infty$ and the limiting distribution of $Z_{t+i\mid t}$ degenerate, and so we rule out this type of unrealistic behaviour for applications.\
    To allow for more flexible scale normings we also consider the following form for the norming functions, which are inspired by Model 3 from \cite{wadstawn22}}
  & \textbf{Model 2:}\quad a_i(x) = \alpha_ix, \quad b_i(x) = 1 + a_i(x)^{\beta},    \label{Mod2Norm}
\end{align}
with the same parameter restrictions as in Model 1.\ In future work
further models for the scale functions $b_i$ may also be considered,
such as Model 1 of \cite{wadstawn22}, but in this paper we will
restrict our attention to models (\ref{Mod1Norm}) and
(\ref{Mod2Norm}).

We will be most interested in cases where there may be assumed to be
some structure in the constants $\bm{\alpha}_{1\,:\,k}$.\ In
particular, when the time series is Markov, as discussed in Section
\ref{Ex}, the $k$ parameters $\bm{\alpha}_{1\,:\,k}$ may be reduced to
a single parameter $\alpha$, greatly simplifying inference.\ In
principle, the constants $\bm{\alpha}_{1\,:\,k}$ should be subject to
constraints in order to have been generated by a stationary process,
i.e., they cannot take arbitrary forms.\ As in \cite{wadstawn22}, we
do not pursue this general question although this would be a useful
avenue for future research in order to be able to propose more
theoretically justified models for non-Markov time series.

\subsection{Semi-parametric modelling}  \label{SemiParamSec}

We first consider a standard method from the conditional extremes
literature \citep{hefftawn04, keefpaptawn13} to construct a composite
log-likelihood for the parameters $\bm{\alpha}_{1\,:\,k}$ and $\beta$
under simplifying assumptions on the structure of
$\bm Z_{t+1\,:\,t+k\mid t}$.\ Specifically, we will make a temporary
working assumption that the copula of $\bm Z_{t+1\,:\,t+k\mid t}$ is
that of independence and that for each $t\in T_u$, the $i$-th
component, $Z_{t+i\mid t},$ of the residual vector
$\bm Z_{t+1\,:\,t+k\mid t}$ has density
\begin{equation}
  f_i(z) = \frac{\delta_i}{2\sigma_i\Gamma(1/\delta_i)}\,\exp\,\left\{-\lvert (z - \mu_i)/\sigma_i\rvert^{\delta_i}\right\}, \quad \mu_i \in \mathbb{R}, \,\, \sigma_i > 0, \,\, \delta_i > 0. \label{DL_density}
\end{equation}
The particular cases where $(\mu_i, \sigma_i, \delta_i) = (0, 1, 2)$
and $(\mu_i, \sigma_i, \delta_i) = (0, 1, 1)$ correspond to the
standard Gaussian and Laplace distributions, respectively.\ Although
in the semi-parametric approach to inference the marginal model for
the residuals is only a temporary working assumption, the inclusion of
the tail parameter $\delta_i$, allows for a more accurate model than
the standard Gaussian assumption, which in turn it is hoped will lead
to better identification of other parameters of interest.\ Random
variables having density function (\ref{DL_density}) appear far back
in the statistical literature \citep{subbotin23, varanasi89} where
they are said to have a generalized Gaussian distribution.\ More
recently they appear in the spatial conditional extremes literature
\citep{wadstawn22, shooter21} and are said to have a $\delta$-Laplace
distribution.\ We will follow this more recent convention, and when a
random variable $Z$ has $\delta$-Laplace distribution with density
(\ref{DL_density}) we write
$Z \sim \delta\text{Laplace}(\mu_i, \sigma_i, \delta_i)$.\ An
alternative working assumption would be to temporarily assume an
independence copula and $Z_{t+i\mid t} \sim N(\mu_i, \sigma^2_i)$ as
is done \cite{hefftawn04}.\ This has the benefit of speeding up
computations and in Section \ref{DataApp} where several models need to
be fit over many bootstrap samples we prefer to work with this
assumption.\ However based on simulations estimating the subasymptotic
extremal index as described in Section \ref{Mark1}, the
$\delta$-Laplace assumption gave slightly better results and so
elsewhere we will use it in what follows.

Under these working assumptions, and writing $\bm{\theta}=(\bm{\alpha}_{1\,:\,k}, \beta, \bm{\mu}_{1\,:\,k}, \bm{\sigma}_{1\,:\,k}, \bm{\delta}_{1\,:\,k})$ we obtain the composite log-likelihood under either Model 1 or 2 normings as 
\begin{equation}
  l(\bm{\theta}) = \sum_{t \in T_u}\sum_{i=1}^{k} \bigg\{\text{log}\,\delta_i - 
  \text{log}\,b_i(x_t) - \text{log}\,\sigma_i - 
  \text{log}\,\Gamma\,(1/ \delta_i) 
  - \left\lvert \frac{x_{t+i} - a_i(x_t) - b_i(x_t)\mu_i}{b_i(x_t)\sigma_i}\right\rvert^{\delta_i} \bigg\}.  \label{Comp1}
\end{equation}
When $|t - t'| < k$ with $t,t'\in T_u $, the residual vectors
$\bm Z_{t+1\,:\,t+k\mid t}$ and $\bm Z_{t'+1\,:\,t'+k \mid t'}$
contain duplicate information, i.e., some values appear in both
vectors.\ This fact together with our false working assumptions
regarding the residual vectors means that $\ell(\bm \theta)$ in
expression (\ref{Comp1}) does not correspond to a proper likelihood
function.\ The use of composite likelihoods is standard in other
applications of the conditional extremes model.\ The parameters of
interest are $\bm{\alpha}_{1\,:\,k}$ and $\beta$ while
$(\bm{\mu}_{1\,:\,k}, \bm{\sigma}_{1\,:\,k}, \bm{\delta}_{1\,:\,k})$
correspond to nuisance parameters.\ We estimate
$(\bm{\alpha}_{1\,:\,k}, \beta)$ by maximizing the corresponding
profile log composite-likelihood
\begin{equation}
l(\bm{\alpha}_{1\,:\,k}, \beta) = \text{sup}_{(\bm{\mu}_{1\,:\,k}, \bm{\sigma}_{1\,:\,k}, \bm{\delta}_{1\,:\,k})}
 l(\bm{\theta}).  \label{PseudoProf}
\end{equation}
For fixed proposed values of $(\bm{\alpha}_{1\,:\,k}, \beta)$ in
(\ref{PseudoProf}) the values of
$(\bm{\mu}_{1\,:\,k}, \bm{\sigma}_{1\,:\,k}, \bm{\delta}_{1\,:\,k})$
maximizing the right-hand side are obtained as the maximum likelihood
estimates of the appropriate $\delta$-Laplace samples.\ Specifically,
for each $i=1{\,:\,}k$, for proposed values of
$(\bm{\alpha}_{1\,:\,k}, \beta)$, we compute the empirical lag $i$
residuals as
\begin{equation}
Z_{t+i\mid t} = \frac{X_{t+i} - a_i(X_t)}{b_i(X_t)}, \quad t \in T_u.  \label{Lag_i_Res}
\end{equation}
Under our working assumptions, the residuals
$\{Z_{t+i\mid t}\}_{t\in T_u}$ are a random sample from a
$\delta$-Laplace distribution with density (\ref{DL_density}).\ The
values $(\mu_i, \sigma_i, \delta_i)$ are obtained as the values
maximizing the $\delta$-Laplace log-likelihood function
$\sum_{t\in\,T_u}\text{log}\,f_i(z_{i\mid t})$ with $f_i$ as in
\eqref{DL_density}).\ Although these do not have a closed form
solution, they are straightforward enough to obtain numerically.

Having obtained point estimates
$(\hat{\bm{\alpha}}_{1\,:\,k}, \hat{\beta} )$ of
$(\bm{\alpha}_{1\,:\,k}, \beta),$ an estimated realization of the
residual vector $\bm Z_{t+1\,:t+k\mid t}$ may be obtained by randomly
sampling from the fitted empirical residuals
\begin{equation}
  \hat{\bm Z}_{t+1\,:\,t+k\mid t} = \frac{\bm{X}_{t+1\,:\,t+k} - \hat{\bm{a}}_{1\,:\,k}(X_t)}{\hat{\bm{b}}_{1\,:\,k}(X_t)}, \quad t \in T_u.  \label{EmpRes}
\end{equation}
If we instead make a working Gaussian assumption
$Z_{t+i\mid t} \sim N(\mu_i, \sigma^2_i)$, then when we profile out
$\mu_i$ and $\sigma_i$ we have closed form solutions for these as the
sample mean and standard deviation of the
$\hat{Z}_{t+i\mid t}, t\in\,T_u$ in (\ref{EmpRes}).\ Thus, in this
case the profile log composite-likelihood is
\begin{equation*}
l(\bm{\alpha}_{1\,:\,k}, \beta) =  \sum_{t \in T_u}\sum_{i=1}^{k} \bigg\{\text{log}\,\hat{\sigma}_i - \text{log}\,b_i(x_t) - \frac{\big(x_{t+i} - a_i(x_t) - b_i(x_t)\hat{\mu}_i\big)^2}{2b_i(x_t)\hat{\sigma}_i} \bigg\},
\end{equation*}
where $\hat{\mu}_i$ and $\hat{\sigma}_i$ are the sample mean and standard deviation of $\{\hat{Z}_{t+i\mid t}\}_{t\in\,T_u}$, respectively.

\subsection{Parametric modelling}
\label{ParamSec}
We now consider
parametric modelling of the residual vector
$\bm Z_{t+1\,:\,t+k\mid t}$.\ We will assume that
$Z_{t+i\mid t} \sim \delta\text{Laplace}(\mu_i, \sigma_i, \delta_i)$
and our interest is in parameterizing the $\delta$-Laplace parameters
as functions of $i$.\ We defer the identification of possible
parametric forms for these parameters until Section \ref{Ex}.\ For
now, we note that the assumption
$Z_{t+i\mid t} \sim \delta\text{Laplace}(\mu_i, \sigma_i, \delta_i)$
implies that
$X_{t+i}~|~X_t > u\sim \delta\text{Laplace}(a_i(X_t) + b_i(X_t)\mu_i,
b_i(X_t)\sigma_i, \delta_i)$.\ For large $i$, we would expect
$X_{t+i}$ to be approximately uncorrelated with $X_t$ so that the
distribution of $X_{t+i}~|~X_t>u$ should be approximately unit
Laplace, i.e., $\delta\text{Laplace}(0,1,1)$, which is the
unconditional distribution of $X_{t+i}$.\ This expectation may be
justified provided that the process satisfies an appropriate mixing
condition which limits long range dependence.\ For the Markov
processes considered in Section \ref{Ex}, several strong mixing
conditions hold \citep[Section 3]{bradley05}.\ In such cases, for
large $i$, we can ensure that the conditional distribution of
$X_{t+i}$ given $X_t > u$ is approximately
$\delta\text{Laplace}(0,1,1)$, in the case of Model 2 normings, by
specifying functions $\mu_i, \sigma_i$ and $\delta_i$ such that
$\mu_i \to 0, \sigma_i\to 1$ and $\delta_i \to 1$ as $i\to \infty$.\
Model 1 requires a slightly more careful parameterization for the
scale parameter $\sigma_i$ and this is discussed further in Section
\ref{Ex}.

We will also assume that the copula of $\bm Z_{t+1\,:\,t+k\mid t}$ is
a Gaussian copula with positive definite correlation matrix $P$.\ The
precise form of $P$ will be context dependent but we will typically
model it as the conditional correlation matrix of a
$(k+1)\mhyphen \textnormal{dimensional}$ random vector conditioned on
the first component being known.\ In many cases, such as the
asymptotically independent Markov processes considered in Section
\ref{Ex}, the $(k+1)\mhyphen \textnormal{dimensional}$ vector may be
naturally taken to be a block of length $k+1$ of a stationary
autoregressive process whose order is equal to that of the underlying
Markov sequence.\ One motivation for the conditional specification of
$P$ can be seen by also including location and scale normalizations
for $X_0$ in (\ref{ForwardAssumption}), i.e., considering limits of
$\{\bm{X}_{0\,:\,{k}} - \bm{a}_{0\,:\,k}(X_0)\}/\bm{b}_{0\,:\,k}(X_0)$
conditioned on $X_0>u$ as $u\to\infty$.  Clearly we may take
$a_0(x)=x$ and $b_0(x)=1$ so that
$\{\bm{X}_{0\,:\,{k}} -
\bm{a}_{0\,:\,k}(X_0)\}/\bm{b}_{0\,:\,k}(X_0)~|~X_0>u\to \bm
Z_{0\,:\,ak}$ where $Z_0=0$ and $\bm Z_{1\,:\,k}$ is as in
(\ref{ForwardAssumption}).\ Thus, our specification of $P$ can be
thought of as specifying the dependence structure of
$\bm Z_{1\,:\,k}~|~Z_0=0$.\ This conditional specification $P$ is
based on an approach in \cite{wadstawn22} where they specify the
dependence structure of a spatial field of residuals conditional on an
extreme observation at a specific location.\ Another motivation for
this approach is that the hidden tail chains in \cite{papaetal17} and
\cite{papastawn20} often have non-stationary dependence structures.\
By modelling $P$ in this conditional manner we may more accurately
model the dependence in the residual vector than by directly
specifying $P$ as the correlation matrix of a block of length $k$ of a
stationary process.

Our assumptions imply that the random vector
$(\Phi^{-1}\{F_{t+1}(Z_{t+1\mid
  t})\},\ldots,\Phi^{-1}\{F_{t+k}(Z_{t+k\mid t})\})$, has a
$k$-dimensional Gaussian distribution with zero mean and correlation
matrix $P$, where $\Phi$ is the standard univariate Gaussian
distribution function and $F_{t+i}$ the distribution function of
$Z_{t+i\mid t}$.\ Although the choice of the Gaussian copula for the
residual vector is not, in general, supported by theory, \cite{towe19}
find that, in a spatial setting, obtaining reliable estimates of the
\cite{hefftawn04} regression parameters $\alpha$ and $\beta$ is more
important than an accurate distributional model for the residuals.\
Moreover, they find that the Gaussian copula model performs well
regardless of whether the process exhibits asymptotic dependence or
independence.\ While, in theory, other copula models could be
considered, computational requirements would then restrict $k$ to
small values.

Subject to our assumptions on the structure of
$\bm Z_{t+1\,:\,t+k\mid t}$, if $\pi_{1\,:\,k\mid t}(\bm{x})$ denotes
the conditional density of $\bm{X}_{t+1\,:\,t+k}$ given $X_t > u$ then
\begin{IEEEeqnarray}{rCl}
  \text{log}\,\pi_{1\,:\,k\mid t}(\bm{x}) &=& \sum_{i=1}^{k}
  \bigg\{\text{log}\,\delta_i - \text{log}\,b_i(x_t) -
  \text{log}\,\sigma_i -
  \text{log}\,\Gamma\,(1/ \delta_i)  \\
  && - \left\lvert \frac{x_{t+i} - a_i(x_t) -
      b_i(x_t)\mu_i}{b_i(x_t)\sigma_i}\right\rvert^{\delta_i} \bigg\}
  \notag +\, 0.5\,\text{log}\,|Q| - 0.5\,\bm{w}^TQ\bm{w} +
  0.5\,\sum_{i=1}^{k}\big[
  \Phi^{-1}\{F_{t+i\mid t}(x_{t+i})\}\big]^2 \label{LogCondDens}
\end{IEEEeqnarray}
where $Q = P^{-1}$,
$\bm{w} = (\Phi^{-1}\{F_{t+1\mid t}(x_{t+1})\},\ldots,
\Phi^{-1}\{F_{t+k\mid t}(x_{t+k})\}) $ with $F_{t+i\mid t}$ the
conditional distribution function of $X_{t+i}$ given $X_t > u$.\ The
composite likelihood is given by
$\sum_{t\in T_u}\text{log}\,\pi_{1\,:\,k\mid t}(\bm{x})$.

\subsection{Quantification of uncertainty and model diagnostics}  \label{DiagSec}

Uncertainty in parameter estimates and other quantities of interest,
such as those discussed in Section \ref{SimulationSec}, are obtained
via bootstrapping methods.\ For stationary time series, standard
methods of obtaining replicate bootstrap samples include the block
bootstrap \citep{carlstein86} and generalizations such as the moving
block bootstrap \citep{kunsch89} and the stationary bootstrap
\citep{polrom94}.\ The block bootstrap requires us to sample with
replacement from non-overlapping blocks
$\bm{X}_{1\,:\,b}, \bm{X}_{b+1\,:\,2b}, \bm{X}_{2b+1\,:3b},\ldots$ of
length $b \in \mathbb{N}$.\ Successively sampled blocks are then
joined together to produce a single bootstrap sample.\ If the length
of the original time series is not a multiple of $b$ then we simply
truncate the last sampled block so that the length of the bootstrap
sample matches that of the original series.\ The moving block
bootstrap generalizes this procedure by sampling with replacement from
the overlapping blocks
$\bm{X}_{1\,:\,b}, \bm{X}_{2\,:\,b+1},\bm{X}_{3\,:\,b+2},\ldots$.\ The
stationary bootstrap further generalizes this procedure by sampling
blocks of random, geometrically distributed, lengths.\ The stationary
bootstrap has some appeal due to the fact that, unlike the block and
moving block bootstraps, it produces samples that are stationary.\
However, theoretical results from \cite{lahiri99} suggest that the use
of non-random blocks leads to lower mean squared errors and moreover,
overlapping blocks are preferable to non-overlapping blocks.\ In our
data application of Section \ref{DataApp} we thus opt for the moving
block bootstrap to quantify uncertainty in parameter estimates and
other quantities of interest.

As our fitted model will be used for extrapolation beyond the fitting
threshold $u$, potentially at thresholds larger than any observation,
it is important that estimates of the parameters $\bm{\alpha}_{1\,:\,k}$
and $\beta$ are stable above $u$.\ This may be checked by fitting our
conditional model above a range of thresholds and assessing
graphically when parameter estimates stabilise.\ In order to maximize
the amount of data available for fitting our model, we take $u$ to be
the smallest threshold above which parameter estimates are stable.\ In
practical examples, $\bm{\alpha}_{1\,:\,k}$ will contain fewer than
$k$ free parameters so that only a few plots need to be inspected.\
Also, a basic modelling assumption is that the residual vector
$\bm Z_{t+1\,:\,t\,:\,k\mid t}$, is conditionally independent of $X_t$
given $X_t>u$.\ The validity of this may be checked, either
informally, via scatter plots of $(X_t, \hat{Z}_{t+i\mid t}), t\in T_u,$ for
a selection of components $i\in 1\,:\,k$, or formally via hypothesis
testing.

One approach to selecting between different models, is to compare
estimates from models with those obtained empirically, of various
cluster functionals.\ Two such functionals that may be used are
\begin{align}
\theta(v, d) &= \mathbb{P}\big(X_2 \leq v,\ldots, X_d \leq v \mid X_1 > v\big),   \label{SAEI}   \\
\chi(v, d)  & = \mathbb{P}(X_{d+1} > v \mid X_1 > v).  \label{SAE}  
\end{align}
The probabilities in (\ref{SAEI}) and (\ref{SAE}) correspond to
subasymptotic versions of the extremal index $\theta$ \citep{lead83}
and the upper tail dependence measure $\chi,$ and are both explored in
\cite{ledtawn03}.\ Both (\ref{SAEI}) and (\ref{SAE}) are used in
\cite{winttawn16} and \cite{Winter2017} for discriminating between
models and as a diagnostic to detect the appropriate order in their
extremal Markov models.\ At moderately high thresholds $v,$ above
which there is a reasonable amount of data, empirical estimates of
(\ref{SAEI}) and (\ref{SAE}) will be quite accurate.\ By inspecting
how empirical estimates vary as $v$ increases and comparing with
estimates obtained by models, we may select models that most closely
match the behaviour of the empirical estimates.

When the parametric approach is taken, several other means of model
discrimination and diagnostics become available.\ For example, in a
spatial setting, \cite{wadstawn22} suggest using the Akaike
information criterion with the composite likelihood in place of the
true likelihood, for selecting models.\ A similar approach is taken in
a Bayesian setting in \cite{shooter19}.\ From a composite likelihood,
a composite posterior is constructed, which is used like the true
posterior for the purposes of calculating the deviance information
criterion to discriminate between models.\ Information criterion
specifically calibrated for composite likelihoods have also been
considered in \cite{varin05} and \cite{Ng14}.

For parametric models, we may separately test the goodness of fit for
the marginal components and copula of the residual vectors
$\bm Z_{t+1\,:\,t+k\mid t}$.\  By computing the empirical lag $i$
residuals
\begin{equation}
Z_{t+i\mid t} = \frac{X_{t+i} - \hat{a}_i(X_t)}{\hat{b}_i(X_t)}, \quad t \in T_u,
\end{equation}
compatibility with the fitted
$\delta\text{Laplace}(\mu_i, \sigma_i, \delta_i)$ distribution may be
assessed via standard methods such as quantile-quantile plots or more
formal hypothesis tests.\ Furthermore, the goodness of fit for the
copula of the residual vectors $\bm Z_{t+1\,:\,t+k\mid t}$ can be
carried out using tests found in \cite{genest09}.

\section{Simulation methods for rare event estimation}
\label{SimulationSec}
We now consider the main purpose of fitting the conditional extremes
models discussed so far: simulation of a stationary time series when
in an extreme state.\ Section \ref{ForSim} discusses simulating
forward in time from an exceedance of a large threshold.\ This may be
used to estimate conditional expectations of the form
$\mathbb{E}(g(\bm{X}_{1:d})~|~ X_1 > v), v \geq u, d \in \mathbb{N},$
for some function of interest $g$.\ It also may be used to simulate
replicate clusters of exceedances from which various functionals of
interest may be calculated.\ Typical examples include the cluster
maxima, mean cluster size or, as considered in \cite{winttawn16}, the
maximum number of consecutive exceedances within a cluster.

Section \ref{ImpSamp} introduces the importance sampling method of
\cite{owen19}.\ This allows for estimation of conditional expectations
of the form
$\mathbb{E}(g(\bm{X}_{1:d})~|~\max\bm{X}_{1:d} > v), v \geq u, d\in
\mathbb{N},$ for some function of interest $g$.\ Here the conditioning
is on there being at least one exceedance anywhere within a block of
observations rather than at the start of the block as is the case in
Section \ref{ForSim}.

It is tacitly assumed in Sections \ref{ForSim} and \ref{ImpSamp} that
$d - 1 \leq k$.\ That is, we do not consider estimation of events that
involve observations at a larger lag from a threshold exceedance than
the block length, $k$, used in fitting our conditional model.\
Although in examples such as those considered in Section \ref{Ex}, we
may allow for $d-1 > k$ by extrapolating the structure in
$\bm{\alpha}_{1\,:\,k}$ to $\alpha_i$, $i > k$, in such cases it may be
preferable to simply select $k$ to be at least as large a lag as we
want to use for simulation purposes.

\subsection{Forward simulation} \label{ForSim}

From a fitted conditional model for $\bm{X}_{t+1\,:\,t+k}$ we may simulate up to $k$-steps forward in time from the extreme event $\{X_t > u\}$ by rearranging (\ref{EmpRes}) to get 
\begin{equation}
\bm{X}_{t+1\,:\,t+k} = \hat{\bm{a}}_{1\,:\,k}(X_t) +  \hat{\bm{b}}_{1\,:\,k}(X_t)\hat{\bm Z}_{t+1\,:\,t+k\mid t} , \quad t\in T_u.
\end{equation}
The residual vector $\hat{\bm Z}_{t+1\,:\,t+k\mid t}$ may be obtained
either empirically as in Section \ref{SemiParamSec} or simulated from
a Gaussian copula model as in Section \ref{ParamSec}.\ Simulating
forward in time from an extreme event allows us to easily estimate
various quantities of interest.\ In addition to $\theta(v, d)$ and
$\chi(v, d) $ defined in (\ref{SAEI}) and (\ref{SAE}), other possible
quantities of interest include
\begin{align}
e_1(v,d) &= \mathbb{E}(\text{max\,}\bm{X}_{1:d}~|~ X_1 > v), \label{ClusterMax} \\
e_2(v,d) &= \mathbb{E}(d^{-1}\sum_{i=1}^{d}X_i~|~ X_1 > v)  \label{ClusterMean} \\
e_3(v,d) &= \mathbb{E}\bigg(\sum_{i=1}^{d}\mathbbm{1}[X_{i} > v]  \Bigm| X_1 > v \bigg)  \label{TotalExc} \\
p(v, d, r) & = \mathbb{P}\bigg(\sum_{i=1}^{d}\mathbbm{1}[X_{i} > v] =r \Bigm| X_1 > v \bigg), \label{CondDist}
\end{align}
where $v > u$.\ Quantities such as (\ref{ClusterMax})-(\ref{TotalExc})
may be used as simple summary statistics to help build up a picture of
how a process behaves after entering an extreme state.\ The
probability in (\ref{CondDist}) concerns the distribution of the
number of exceedances in a block of size $d$ given an exceedance at
the start of the block.\ All of these quantities are of the form
$\mathbb{E}(g(\bm{X}_{1:d})~|~ X_1 > v)$ for a suitable choice of
function $g$.\ For example, for (\ref{CondDist}) we have
$g(\bm{x}) = \mathbbm{1}\big[\sum_{i=1}^{d}\mathbbm{1}[x_i > v]
=r\big]$.

As the distribution of $(X_t - u)~|~X_t > u$ is unit exponential, due
to the assumed Laplace margins, we may estimate
$\mathbb{E}(g(\bm{X}_{1:d})~|~ X_1 > v)$ using Algorithm
\ref{SAEI_Alg}.\ This involves repeatedly simulating forward from the
exceedance $X_1$ and estimating the quantity of interest via an
empirical proportion.\ Exactly how step 4 is carried out will depend
on whether semi-parametric inference or parametric inference is used.\
For semi-parametric inference, step 4 involves randomly sampling from
the empirical residuals (\ref{EmpRes}) whereas for parametric
modelling it will involve simulation from a Gaussian copula.

We may also wish to estimate the cluster size distribution.\ Suppose
that clusters are defined by the runs method \citep{smithweiss94} with
run length $r$.\ Thus, a cluster will be said to be initialized when a
threshold $v$ is exceeded and ends when $r$ consecutive
non-exceedances occur.\ Using this definition, clusters are simulated
in \cite{Winter2017} by iteratively simulating forward from the first
cluster exceedance until $r$ consecutive non-exceedances occur.\ In
our case, we simulate jointly a full block of length $k$ forward from
the first exceedance, where the first exceedance is simulated by
setting $X_t = v + E$ where $E\sim \exp(1)$.\ The constant $k$
should be chosen so that the probability of observing a cluster of
length greater than $k$ is negligible.\ The simulated block
$\bm{X}_{t:t+k}$ will then typically contain several values after the
cluster has terminated and we may then retain the smaller block
$\bm{X}_{t:t+k'},$ where $k' < k,$ corresponding to a single simulated
cluster.
\begin{algorithm}
  \caption{Estimation of $\mathbb{E}(g(\bm{X}_{1\,:\,d})~|~ X_1 > v)$
    via forward simulation.}
\SetKwData{Left}{left}\SetKwData{This}{this}\SetKwData{Up}{up}
\SetKwFunction{Union}{Union}\SetKwFunction{FindCompress}{FindCompress}
\SetKwInOut{Input}{input}\SetKwInOut{Output}{output}

\Input{Threshold $v > u, d, n \in \mathbb{N}$ and constants
  $(\hat{\bm{\alpha}}_{1\,:\,k}, \hat{\beta})$ from fitted conditional
  model.}
\Output{An estimate of
  $\mathbb{E}(g(\bm{X}_{1\,:\,d})~|~ X_1 > v)$} \For{$i\leftarrow 1$
  \KwTo $n$}{ \emph{simulate exceedance amount} $E \sim$ $\exp(1)$ \;
  \emph{set} $X^i_1 = v + E$ \; \emph{simulate residual}
  $\hat{\bm Z}^{(1), i}_{2\,:\,d}$ \emph{from fitted conditional model
    independently of $X_1$} \; \emph{set}
  $\bm{X}^{i}_{2\,:\,d} = \hat{\bm{\alpha}}_{1\,:\,d-1}X^i_1 +
  (X^i_1)^{\hat{\beta}}\hat{\bm Z}^{(1), i}_{2\,:\,d}$ \; \emph{set}
  $\bm{X}^{i}_{1\,:\,d} = (X_1^i, \bm{X}^{i}_{2\,:\,d})$ \; }
\emph{return}
$\hat{\mathbb{E}}(g(\bm{X}_{1\,:\,d})~|~ X_1 > v) = n^{-1}\sum_{i=1}^n
g(\bm{X}^{i}_{1\,:\,d})$.
\label{SAEI_Alg}
\end{algorithm} 
\subsection{Importance sampling}  \label{ImpSamp}

In this section we introduce the estimator of \cite{owen19} for
estimating probabilities of the form
$\mathbb{P}(\cup_{i=1}^{d} \{X_i > v_i\}),$ where
$v_i \in (0, \infty)$, $i\in\,1{\,:\,}d$.\ The same approach can be
found in the simulation algorithms of \cite{wadstawn22} and
\cite{adler12} in slightly different settings.\ Although the estimator
may be used for estimating the probability of a union of arbitrary
events, we restrict attention to events of the specific form
$\{X_i > v_i\}$ as these are of most interest to us in our time series
context.\ Moreover, we will see how the estimator may be adapted to
allow us to estimate the probabilities of other events of interest.

Consider a block $\bm{X}_{1\,:\,d}$ of length $d$ from a time series
with joint density function $\pi$.\ For $i\in\,1{\,:\,}d,$ let
$L_i \subseteq \mathbb{R}^d$ be the region
$L_i = \{\bm{x} \in \mathbb{R}^d : x_i > v_i\}$ and let
$\mathcal{L} = \cup_{i=1}^d L_i$.\ We will be most interested in the
case where each $v_i, 1 \leq i \leq d$, is a large quantile of the
standard Laplace distribution, so that if $\bm{X}_{1\,:\,d}$ lies in
$\mathcal{L},$ then at least one of its components is large.\ We
consider estimation of
\begin{equation}
p = \mathbb{P}\bigg(\bigcup_{i=1}^{d} \{X_i > v_i\}\bigg) = \mathbb{P}(\bm{X}_{1\,:\,d} \in \mathcal{L}).
\end{equation}
A special case frequently of interest is when all the thresholds are
equal, say $v_i = v, 1\leq i \leq d$, in which case
$p = \mathbb{P}(\text{max\,}\bm{X}_{1\,:\,d} > v)$ is the probability
of exceeding the threshold $v$ within a block of $d$ observations.\
The obvious empirical estimator,
$n^{-1}\sum_{i=1}^n \mathbbm{1}_{\mathcal{L}}(\bm{X}^{i}_{1\,:\,d}),$
of $p$ based on $n$ independent replications,
$\{\bm{X}^{i}_{1\,:\,d}\}_{i=1}^{n}$, of $\bm{X}_{1\,:\,d},$ is
unbiased and has variance $p(1-p)/n$.\ \cite{owen19} show that this
estimator may be improved upon, in the sense of reduced variance, by
sampling from an appropriate mixture distribution instead of directly
from $\bm{X}_{1\,:\,d}$.\ Specifically, for each $i \in 1{\,:\,}d,$
define $\pi_i^{*}$ to be the conditional density of $\bm{X}_{1\,:\,d}$
given $X_i > v_i$, so that
$\pi_i^{*}(\bm{x}) = \pi(\bm{x})\mathbbm{1}_{L_i}(\bm{x})/p_i,
\bm{x}\in\mathbb{R}^d,$ where $p_i = \mathbb{P}(X_i > v_i) > 0$.\ The
importance sampling density proposed by \cite{owen19} is
$\pi^{*} = \sum_{i=1}^{d}w_i\pi_i^{*}$ where $w_i = p_i / \bar{p}$ and
$\bar{p} = \sum_{i=1}^{d}p_i$ is the union bound of $p$.\ Thus the
mixture component $\pi_i^{*}$ is sampled from with probability
proportional to $p_i$.\ Since
\begin{equation}
p = \mathbb{E}_{\pi}\{\mathbbm{1}_{\mathcal{L}}(\bm{X}_{1\,:\,d})\} = \mathbb{E}_{\pi^{*}}\bigg\{\frac{\mathbbm{1}_{\mathcal{L}}(\bm{X}_{1\,:\,d})\pi(\bm{X}_{1\,:\,d})}{\pi^{*}(\bm{X}_{1\,:\,d})} \bigg\},
\end{equation}
this motivates the following estimator of $p$ 
\begin{equation}
  \hat{p} = 
  \frac{1}{n}\sum_{i=1}^{n} \frac{\mathbbm{1}_{\mathcal{L}}(\bm{X}^{i}_{1\,:\,d})\pi(\bm{X}^{i}_{1\,:\,d})}{\pi^{*}(\bm{X}^{i}_{1\,:\,d})}  =
  \frac{1}{n}\sum_{i=1}^{n} \frac{\mathbbm{1}_{\mathcal{L}}(\bm{X}^{i}_{1\,:\,d})\pi(\bm{X}^{i}_{1\,:\,d})}{\sum_{j=1}^{d} \mathbbm{1}_{L_j}(\bm{X}^{i}_{1\,:\,d})\pi(\bm{X}^{i}_{1\,:\,d})\bar{p}^{-1}}, \quad \bm{X}^{i}_{1\,:\,d} \overset{iid}{\sim} \pi^{*}. \label{owens1}
\end{equation}
As $\mathbbm{1}_{\mathcal{L}}(\bm{X}^{i}_{1\,:\,d}) = 1$ when $\bm{X}^{i}_{1\,:\,d}\sim \pi^{*}$, estimator (\ref{owens1}) simplifies to 
\begin{equation}
\hat{p} = \frac{\bar{p}}{n}\sum_{i=1}^{n}\frac{1}{S(\bm{X}^{i}_{1\,:\,d})}, \quad \bm{X}^{i}_{1\,:\,d} \overset{iid}{\sim} \pi^{*},  \label{owens2}
\end{equation}
where
$S(\bm{X}^{i}_{1\,:\,d}) =
\sum_{j=1}^{d}\mathbbm{1}_{L_j}(\bm{X}^{i}_{1\,:\,d})$ counts the
number of events $\{X_i > v_i\}, 1 < i < d$, that occur in the block
$\bm{X}^{i}_{1\,:\,d}$ of length $d$.\ As
$1 \leq S(\bm{X}^{i}_{1\,:\,d}) \leq d$ when
$\bm{X}^{i}_{1\,:\,d} \sim \pi^{*}$, $\hat{p}$ is always well defined
and respects the theoretical bounds $\bar{p}/d \leq p \leq \bar{p}$.

The union bound $\bar{p}$ which appears in expression (\ref{owens2})
is easily obtained using the assumption that the margins of the
process are standard Laplace distributed.\ For the rest of this
section we focus on the case of a common threshold
$v_i = v, i \in 1{\,:\,}d,$ in which case $\bar{p} = de^{-v}/2$.\ In
order to be able to estimate $p$ via (\ref{owens2}), we need to be
able to simulate repeatedly from the block $\bm{X}_{1\,:\,d}$
conditional on there being at least one exceedance of $v$ within the
block and then count the total number of exceedances.\ As the marginal
distributions are equal and we are assuming a common threshold, we
have $w_i =d^{-1}, i \in 1{\,:\,}d,$ and the distribution $\pi^*_i$ is
the conditional distribution of $\bm{X}_{1\,:\,d}~|~X_i > v$.\ When
$2\leq i \leq d-1,$ we then need to be able to simulate from
$\bm{X}_{1\,:\,i-1}~|~X_i>v$ and $\bm{X}_{i+1\,:\,d}~|~X_i>v,$ i.e.,
we need to be able to simulate both forward and backward in time from
the event $X_i > v$.\ \cite{janseg13} consider the behaviour of vector
valued Markov time series both forward and backward in time from an
extreme event through their so-called forward and backward tail
chains.\ In this context, an extreme corresponds to a large value of
the Euclidean norm, and it is shown that the forward tail chain
determines the backward tail chain and conversely.\ However their
results do not directly apply to our setting due to their multivariate
regular variation \citep{resn87} assumption which excludes
asymptotically independent processes that are not independent.

To simulate backwards in time from a threshold exceedance, from (\ref{ForwardAssumption}), there exist location and scale norming functions
$\{a_i: \mathbb{R} \to \mathbb{R}\}_{i=-k}^{k}$ and
$\{b_i: \mathbb{R} \to \mathbb{R_+}\}_{i=-k}^{k}$ respectively, 
such that for any $k\in\mathbb{N}$
\begin{equation}
\left. \bigg(X_0 - u, \frac{\bm{X}_{-k\,:\,k} - \bm{a}_{-k\,:\,k}(X_0)}{\bm{b}_{-k\,:\,k}(X_0)}\bigg) \right\vert X_0> u \xrightarrow[]{D} (E_0, \bm Z_{-k\,:\,k}),  \label{BackwardAssumption}
\end{equation}
as $u\to\infty,$ where $E_0$ is a unit exponential random variable
independent of the random vector $\bm Z_A \sim G_A,$ $A = -k{\,:\,}k$,
where $G_A$ is a joint distribution function on $\mathbb{R}^{2k}$ with
non-degenerate marginal distributions $G_i$, $i\neq 0$, that place no
mass at $+\infty$.\ In (\ref{BackwardAssumption}) we adopt the
convention that $a_0(x)=x, b_0(x)=1$ and $Z_0 = 0$.\ We may then fit
models for the block $\bm{X}_{t-k\,:\,t+k}$, $k$-steps prior to and
following the exceedance $X_t > u$, using the methods of Section
\ref{StatInf}.\ Thus for example, in the case of Model 1 normings
(\ref{Mod1Norm}), we may fit a model of the form
\begin{equation}
\bm{X}_{t-k\,:\,t+k} \overset{D}{=} \bm{\alpha}_{-k\,:\,k}X_t + X_t^{\bm{\beta}_{-k\,:\,k}}\bm Z_{t-k\,:\,t+k\mid t}, \quad t\in T_u. \label{ForwardBackMod}
\end{equation}
where
$\bm{\beta}_{-k\,:\,k} =(\beta^{-},\ldots,\beta^{-},0,\beta^+,\ldots,
\beta^+),$ with $\beta^-$ and $\beta^+$ scale parameters associated to
observations prior to and following the threshold exceedance at
$X_t$.\ In the case of semi-parametric modelling with parameters
$\bm{\alpha}_{-k\,:\,k}, \beta^{-}$ and $\beta^+$ to estimate, where
$\alpha_0 = 1$, fitting the model (\ref{ForwardBackMod}) requires no
new innovations relative to the methods of Section
\ref{SemiParamSec}.\ As before, we make the working assumption that
$Z_{t+i\mid t} \sim \delta\text{Laplace}(\mu_i,\sigma_i, \delta_i),$
where now $i \in{-k\,:\,k}\backslash\{0\}$.\ The residual vector
$\bm Z_{t-k\,:\,t+k\mid t}$ can be simulated empirically by
rearranging (\ref{ForwardBackMod}) and replacing parameters by their
estimated values.

It is of interest to know whether there is any connection between the
parameters associated to the forward and backward chains, i.e.,
$(\bm{\alpha}_{-k\,:\,-1}, \beta^-)$ and
$(\bm{\alpha}_{1\,:\,k}, \beta^+),$ since in such cases we may be able
to reduce the number of parameters to be estimated and hence improve
efficiency.\ For example, for asymptotically dependent processes one
has $\alpha_{-i} = \alpha_i = 1$ and $\beta^-=\beta^+ = 0$.\ For the
asymptotically independent Markov processes that we consider in
Section \ref{Mark1}, with Gaussian and inverted logistic copulas, one
also has $\alpha_{-i} = \alpha_i$ and $\beta^- = \beta^+$.\ The
analogue of such symmetry in a spatial context is isotropy, and for
first-order Markov time series amounts to exchangeability of the
copula of $(X_t, X_{t+1})$.\ In practice, whether or not
such symmetry exists will be unknown, however, we can always fit both
symmetric and asymmetric models and compare fits using diagnostics
such as those described in Section \ref{DiagSec}.

Having estimated $\bm{\alpha}_{-k\,:\,k}, \beta^{-}$ and $\beta^+$, it
is straightforward to estimate $\mathbb{P}(\max \bm{X}_{1\,:\,d} > v)$
using Algorithm \ref{ALOE}, which assumes that Model 1 normings have
been used and is trivially modified for Model 2 normings.\ Also,
exactly how the residual vectors are simulated will depend on whether
the residuals are modelled non-parametrically as in Section
\ref{SemiParamSec} or parametrically as in Section \ref{ParamSec}.\ 
\cite{owen19} prove that $\hat{p}$ is an unbiased estimator of $p$ and
$\textnormal{var}(\hat{p}) \leq p(\bar{p} - p)/n$, from which it
follows that $\hat{p}$ is a consistent estimator of $p$.\ Although
$\hat{p}$ may be used to estimate the probability of an arbitrary
union of events, it is in the rare event setting, when $p$ and
$\bar{p}$ are small, that it is most efficient since then
$p(\bar{p} - p)$ may be orders of magnitude smaller than $p(1 - p)$.\
Thus, in the rare event setting we increase the precision in
estimation by sampling from $\pi^*$ rather than $\pi$.

We now consider estimating the probability of sub-events of
$\cup_{i=1}^{d}\{X_i >v \}$.\ Typical examples of interest include
$\mathbb{P}(\sum_{i=1}^d \mathbbm{1}[X_i > v] = r)$ or
$\mathbb{P}(\sum_{i=1}^d \mathbbm{1}[X_i > v] \geq r), r \geq 1,$
which correspond to ``exactly $r$ exceedances of'' and ``at least $r$
exceedances of'' the threshold $v$ in the block $\bm{X}_{1\,:\,d}$ of
$d$ observations.\ Let $g$ be a function supported on $\mathcal{L}$,
so that $g(\bm{x}) =0$ for $\bm{x} \in \mathcal{L}^c$.\ We consider
estimation of
$ \mathbb{E}_{\pi}\{g(\bm{X}_{1\,:\,d})\} = \int_{\mathcal{L}}
g(\bm{x})\pi(\bm{x})\,d\bm{x}$.\ We may estimate
$\mathbb{E}_{\pi}\{g(\bm{X}_{1\,:\,d})\}$ via importance sampling from
$\pi^{*}$ as
\begin{equation}
\widehat{\mathbb{E}}_{\pi}\{g(\bm{X}_{1\,:\,d})\} =\frac{\bar{p}}{n}\sum_{i=1}^{n}\frac{g(\bm{X}^{i}_{1\,:\,d})}{S(\bm{X}^{i}_{1\,:\,d})}, \quad \bm{X}^{i}_{1\,:\,d} \overset{iid}{\sim} \pi^{*}. \label{owens3}
\end{equation}
Provided $g$ is bounded, as is the case when $g$ is an indicator
function, (\ref{owens3}) defines a consistent estimator of
$\mathbb{E}_{\pi}\{g(\bm{X}_{1\,:\,d})\}$.\ This is the main content
of Theorem \ref{owen_cor} below which is proved in Appendix
\ref{AppendixProof}.
\begin{theorem} \label{owen_cor}
If $\widehat{\mathbb{E}}_{\pi}\{g(\bm{X}_{1\,:\,d})\}$ is as in (\ref{owens3}) and $g$ is supported on $\mathcal{L}$ then 
\begin{align}
  \mathbb{E}_{\pi^*}\big(\widehat{\mathbb{E}}_{\pi}\{g(\bm{X}_{1\,:\,d})\}\big) &= \mathbb{E}_{\pi}\{g(\bm{X}_{1\,:\,d})\} \\
  \intertext{and}
  \quad \textsf{\normalfont var}\left(\widehat{\mathbb{E}}_{\pi}\{g(\bm{X}_{1\,:\,d})\}\right) &= n^{-1}\bigg(\bar{p}\int_{\mathcal{L}} [\{g(\bm{x})^2\pi(\bm{x})\}/S(\bm{x})]d\bm{x} - \big(\mathbb{E}_{\pi}\{g(\bm{X}_{1\,:\,d})\}\big)^2 \bigg).
\end{align} 
Consequently, if $g$ is a bounded function then
$\widehat{\mathbb{E}}_{\pi}\{g(\bm{X}_{1\,:\,d})\}$ is a consistent
estimator of $\mathbb{E}_{\pi}\{g(\bm{X}_{1\,:\,d})\}$.\ Moreover, if
$g$ is an indicator function, i.e., $g(\bm{x}) \in \{0,1\}$, for all
$\bm{x} \in \mathbb{R}^d,$ then
\begin{equation} \label{UpperBound_Cond} \textsf{\normalfont
    var}\big(\widehat{\mathbb{E}}_{\pi}\{g(\bm{X}_{1\,:\,d})\}\big)
  \leq n^{-1}\mathbb{E}_{\pi}\{g(\bm{X}_{1\,:\,d})\}\big(\bar{p} -
  \mathbb{E}_{\pi}\{g(\bm{X}_{1\,:\,d})\}\big).
\end{equation}
\end{theorem}

\begin{algorithm}[htbp!]
\SetKwData{Left}{left}\SetKwData{This}{this}\SetKwData{Up}{up}
\SetKwFunction{Union}{Union}\SetKwFunction{FindCompress}{FindCompress}
\SetKwInOut{Input}{input}\SetKwInOut{Output}{output}

\Input{Threshold $v > u, n \in \mathbb{N}$ and constants $(\hat{\bm{\alpha}}_{-k\,:\,k},\hat{\beta}^{+}, \hat{\beta}^{-})$ from fitted model (\ref{ForwardBackMod}).}
\Output{An estimate of $\mathbb{P}(\cup_{i=1}^d\{X_ i > v\})$}
\For{$i\leftarrow 1$ \KwTo $n$}{
\emph{sample exceedance time} $j \in \{1,\ldots, d\}$  \emph{uniformly at random}\;
\emph{simulate exceedance amount} $E \sim \exp(1)$ \;
\emph{set} $X^i_j = v + E$ \;
\emph{simulate the residual vector} $\hat{\bm Z}^{(j),i}_{1\,:\,d}$ \;
    \uIf{$j=1$}
    {
        \emph{set} $\bm{X}^i_{2\,:\,d} = \hat{\bm{\alpha}}_{1\,:\,(d-1)}X^i_1 + (X^i_1)^{\hat{\beta}^+}\hat{\bm Z}^{(1), i}_{2\,:\,d}$ \;
        \emph{set} $\bm{X}^{i}_{1\,:\,d} = (X^i_1, \bm{X}^i_{2\,:\,d})$
    }
\uElseIf{$j=d$}
{ 
\emph{set} $\bm{X}^i_{1\,:\,{d-1}} = \hat{\bm{\alpha}}_{-(d-1)\,:\,-1}X^i_d + (X^i_d)^{\hat{\beta}^{-}}\hat{\bm Z}^{(d), i}_{1\,:\,d-1}$ \;
 \emph{set} $\bm{X}^{i}_{1\,:\,d} = (\bm{X}^i_{1\,:\,d-1}, X^i_d)  $   \;
}
\Else{
   \emph{set}  $\bm{X}^i_{1\,:\,{j-1}} = \hat{\bm{\alpha}}_{-(j-1)\,:\,-1}X^i_j + (X^i_j)^{\hat{\beta}^{-}}\hat{\bm Z}^{(j), i}_{1\,:\,j-1}$  \; 
   \emph{set} $\bm{X}^i_{j+1\,:\,d} = \hat{\bm{\alpha}}_{1\,:\,d-j}X^i_j + (X^i_1)^{\hat{\beta}^+}\hat{\bm Z}^{(j), i}_{j+1\,:\,d}$ \;
   \emph{set} $\bm{X}^{i}_{1\,:\,d} = (\bm{X}^i_{1\,:\,{j-1}}, X^i_j,  \bm{X}^i_{j+1\,:\,d})$ \;
  }
\emph{calculate} $S(\bm{X}^{i}_{1\,:\,d}) = \sum_{k=1}^{d} \mathbbm{1}[X^i_k > v]$\;
}
\emph{return} $\hat{p} = \frac{de^{-v}}{2}\frac{1}{n}\sum_{i=1}^{n}\frac{1}{S(\bm{X}^{i}_{1\,:\,d})}$.
\caption{Estimation of the probability of at least one exceedance of
  the threshold $v$ in a block of length $d$ in a stationary time
  series in Laplace margins.}\label{ALOE}
\end{algorithm}

We finally consider the case where we want to estimate an expectation
conditionally on $\cup_{i=1}^{d}\{X_i >v \}$ occurring.\ Let
$\pi^{**}$ be the conditional density of $\bm{X}_{1\,:\,d}$ given
$\cup_{i=1}^{d}\{X_i >v \}$, so that
$\pi^{**}(\bm{x}) =
\pi(\bm{x})\mathbbm{1}_{\mathcal{L}}(\bm{x})p^{-1}$ where
$p = \mathbb{P}(\cup_{i=1}^{d}\{X_i >v \})$.\ We wish to estimate
\begin{equation}
 \mathbb{E}_{\pi^{**}}\{g(\bm{X}_{1\,:\,d})\} = \int g(\bm{x})\pi(\bm{x})\mathbbm{1}_{\mathcal{L}}(\bm{x})p^{-1}d\bm{x} =
 p^{-1}\mathbb{E}_{\pi}\{\mathbbm{1}_{\mathcal{L}}(\bm{X}_{1\,:\,d})g(\bm{X}_{1\,:\,d})\}.  \label{owens4}
\end{equation}
Now, since
$\mathbbm{1}_{\mathcal{L}}(\bm{X}_{1\,:\,d})g(\bm{X}_{1\,:\,d})$ is
supported on $\mathcal{L}$, we may estimate
$\mathbb{E}_{\pi}\{\mathbbm{1}_{\mathcal{L}}(\bm{X}_{1\,:\,d})g(\bm{X}_{1\,:\,d})\}$
using the estimator in (\ref{owens3}) and estimate $p^{-1}$ using the
reciprocal of (\ref{owens2}).\ Thus, from (\ref{owens4}) we may
estimate $\mathbb{E}_{\pi^{**}}\{g(\bm{X}_{1\,:\,d})\}$ using
\begin{equation}
\widehat{\mathbb{E}}_{\pi^{**}}\{g(\bm{X}_{1\,:\,d})\} = \frac{ \sum_{i=1}^{n}g(\bm{X}^{i}_{1\,:\,d}) / S(\bm{X}^{i}_{1\,:\,d})}{\sum_{i=1}^{n}1 / S(\bm{X}^{i}_{1\,:\,d})}, \quad  \bm{X}^{i}_{1\,:\,d} \overset{iid}{\sim} \pi^{*}  \label{owens5}
\end{equation}
which, in the case that $g$ is a bounded function, is a consistent
estimator as it is formed from a ratio of consistent estimators.\ The
estimator (\ref{owens5}) appears in Algorithm 3 of \cite{wadstawn22}.\
A typical example where we may use (\ref{owens5}) is to estimate
\begin{equation}
  p^{*}(v, d, r)  = \mathbb{P}\bigg(\sum_{i=1}^{d}\mathbbm{1}[X_i > v] =r \Bigm| \max\bm{X}_{1\,:\,d} > v \bigg),  \quad 1\leq r \leq d,   \label{CondDist2}
\end{equation}
which as $r$ varies from 1 to $d$, estimates the distribution of the
number of exceedances of the threshold $v$ within the block
$\bm{X}_{1\,:\,d}$ of $d$ observations given at least one exceedance.\
To estimate (\ref{CondDist2}) for fixed $r$, we take $g$ to be the
indicator function
$g(\bm{x}) = \mathbbm{1}\big[\sum_{i=1}^{d}\mathbbm{1}[x_i > v]
=r\big]$.\ Algorithm \ref{ALOE} is easily adapted to estimate
(\ref{CondDist2}) or indeed, more generally, (\ref{owens4}), for an
arbitrary $g$ using the estimator (\ref{owens5}).\ The only amendments
required are on line 19, where in addition to calculating
$S(\bm{X}^{i}_{1\,:\,d}),$ we also calculate
$g(\bm{X}^{i}_{1\,:\,d}),$ and then on line 21 we return the value of
(\ref{owens5}).

\section{Examples and simulation study}  \label{Ex}

\subsection{Norming functions for asymptotically independent Markov time series} \label{Mark1}

Results from \cite{papaetal17} imply that for first-order Markov time
series we may consider a greatly simplified structure to the vector of
constants $\bm{\alpha}_{1\,:\,k}$ compared to the general form given
in Section \ref{StatInf}.\ In particular, if we write,
$\alpha_1 = \alpha$ and then we may take $\alpha_i = \alpha^i$ for
$i \in 1{\,:\,}k$.\ Thus the conditional distribution of
$\bm{X}_{t+1\,:\,t+k}$ given $X_t > u$ has representations under
Models 1 and 2 as
\begin{align}
  \bm{X}_{t+1\,:\,t+k} \mid X_t > u & \overset{D}{=} \alpha^{1\,:\,k}X_t + X_t^{\beta}\bm Z_{t+1\,:\,t+k\mid t}, \quad\quad (\text{Model 1}), \\
  \intertext{and}
  \bm{X}_{t+1\,:\,t+k} \mid X_t > u & \overset{D}{=} \alpha^{1\,:\,k}X_t + \{1 + (\alpha^{1\,:\,k}X_t)^{\beta}\}\bm Z_{t+1\,:\,t+k\mid t}, \quad\quad (\text{Model 2}).
\end{align}
For semi-parametric modelling as in Section \ref{SemiParamSec}, we
then have only two parameters, $\alpha$ and $\beta,$ to estimate.\ The
constants $\bm{\alpha}_{1\,:\,k} = \alpha^{1\,:\,k}$ may also be
defined recursively as $\alpha_t = a(\alpha_{t-1}), 1\leq t \leq k$,
with initial condition $\alpha_0 =1,$ where
$a:\mathbb{R}\to\mathbb{R}$ is the function defined by
$a(x) = \alpha\,x$.\ Moreover, the values $\alpha^{1\,:\,k}$ may be
recognized as the values at lags 1 up to $k$ of the autocorrelation
function of the first-order autoregressive model (\ref{GaussAR1}) with
$\rho = \alpha$.

\cite{papastawn20} suggest that for higher-order Markov sequences,
structure on the constants $\bm{\alpha}_{1\,:\,k}$ may be obtained in
a similar way.\ In particular, for a Markov sequence of order $l< k$,
then given initial values $\alpha_{0:l-1}$ where $\alpha_0 = 1$, we
may obtain $\bm{\alpha}_{l\,:\,k}$ via the recurrence
$\alpha_t = a(\bm{\alpha}_{t-l:t-1})$ where
$a:\mathbb{R}^l\to\mathbb{R}$ is a differentiable and homogenous
function of order 1, i.e.,
$a(t\bm{x}) = ta(\bm{x}), \bm{x} \in \mathbb{R}^l$.\ We will consider
two possible functional forms for the function $a$.\ In the first case
we will assume that $a$ takes the form of an autocorrelation function
of a stationary autoregressive model of order $l$, and the second case
is based on a functional form in \cite{papastawn20}.

For now we focus on the case where our time series is Markov of order
$l=2$.\ Consider the autocorrelation function of the second order
stationary autoregressive model
$Y_{n} = \theta_1Y_{n-1} + \theta_2Y_{n-2} + w_n$ where $\{w_n\}$ is a
zero mean uncorrelated sequence independent of $\{Y_n\}$.\ If $\rho_n$
denotes the value of the autocorrelation function of the sequence
$\{Y_n\}_{n=0}^{\infty}$ at lag $n$, then $\rho_n$ is determined for
all $n$ by the recurrence
$\rho_n = \theta_1\rho_{n-1} + \theta_2\rho_{n-2}, n \geq 2,$ with
initial conditions $\rho_0 = 1, \rho_1 = \theta_1 / (1 - \theta_2)$.\
We use this form of recurrence to define structure on the sequence
$\bm{\alpha}_{1\,:\,k}$ as
\begin{equation}
 \alpha_t = \theta_1\alpha_{t-1} + \theta_2\alpha_{t-2},\quad 2 \leq t \leq k, \quad \text{with \,\,} \alpha_0 = 1, \alpha_1 = \theta_1 / (1 - \theta_2).  \label{Cor2}
\end{equation}
The recurrence in (\ref{Cor2}) may be written alternatively as
$\alpha_t = a(\alpha_{t-1}, \alpha_{t-2})$ where
$a:\mathbb{R}^2\to\mathbb{R}$ is the function
$a(x_1,x_2) = \theta_1x_1 + \theta_2x_2$.\ For this model, in the case
of semi-parametric inference as in Section \ref{SemiParamSec}, the
$k+1$ parameters $(\bm{\alpha}_{1\,:\,k}, \beta)$ to estimate in
(\ref{Mod1Norm}) and (\ref{Mod2Norm}) are reduced to the three
parameters $(\theta_1, \theta_2, \beta)$.\

A difficulty arises when implementing this model due to the fact that,
in order for $(\theta_1, \theta_2)$ to define the autocorrelation
function of a stationary process, they are subject to certain
constraints.\ These constraints are that $(\theta_1, \theta_2)$ lie in
the interior of the triangular region defined by the inequalities
$\theta_2 < 1 + \theta_1$, $\theta_2 < 1 - \theta_1$ and
$\theta_2 > -1$.\ To deal with these constraints we consider a
reparameterization in terms of partial autocorrelations.\
\cite{barn73} showed that a stationary autoregressive process of order
$l$ may be parameterized in terms of the first $l$ partial
autocorrelations which each may be taken to vary freely in $(-1, 1)$.\
Moreover, the partial autocorrelations are shown to be in a
one-to-one, continuously differentiable correspondence with the
autoregression parameters.\ This greatly simplifies inference in
comparison to working directly with the autoregression parameters,
especially as $l$ increases and the parameter constraints become more
complex.\ In the case of an order $2$ process, if $(r_1, r_2)$ are the
first two partial autocorrelations, then the correspondence between
$(r_1, r_2)$ and $(\theta_1, \theta_2)$ is given by
\begin{align}
\theta_1 &= r_1(1 - r_2)   \label{PACF1}  \\
\theta_2 &= r_2   \label{PACF2}
\end{align}
with $(r_1, r_2) \in (-1, 1)^2$.\ Thus, in this parameterization, we
have parameters $(r_1, r_2, \beta)$ to infer.\ From our fitted values
of $r_1$ and $r_2$, we obtain the fitted values of $\theta_1$ and
$\theta_2$ from equations (\ref{PACF1}) and (\ref{PACF2}), and
consequently $\bm{\alpha}_{1\,:\,k}$ from equation (\ref{Cor2}).

This approach generalizes to higher-order cases.\ For example, in the
case of an order 3 Markov sequence, structure on
$\bm{\alpha}_{1\,:\,k}$ comes from the autocorrelation function of an
autoregressive model of order 3 as
\begin{align}
 \alpha_t &= \theta_1\alpha_{t-1} + \theta_2\alpha_{t-2}  +  \theta_3\alpha_{t-3}, \quad 3 \leq t \leq k, \quad 
 \intertext{with}
 \alpha_0 &= 1,  \,\,\, \alpha_1 = \frac{\theta_1 + \theta_2\theta_3}{1 - \theta_2 - \theta_1\theta_3 - \theta_3^2},\,\,\,
 \alpha_2 =  \theta_2 + (\theta_1 + \theta_3)\alpha_1.  \label{Cor3}
\end{align}
The parameters $(\theta_1, \theta_2, \theta_3)$ are subject to the
stationarity constraints
$\theta_1 + \theta_2 + \theta_3 < 1$, $-\theta_1 + \theta_2 - \theta_3 < 1$, $\theta_3(\theta_3 - \theta_1) - \theta_2 < 1$ and $|\theta_3| < 1$.\\
We then reparameterize in terms of $(r_1, r_2, r_3)$ where
\begin{align*}
\theta_1 &= r_1 - r_1r_2 - r_2r_3 \\
\theta_2 &= r_2 - r_1r_3 + r_1r_2r_3 \\
\theta_3 &= r_3
\end{align*}
where $(r_1, r_2, r_3) \in (-1,1)^3$.\ Although in principle this
approach may used for Markov processes of any order $l$, the initial
conditions for $\bm{\alpha}_{0\,:\,l-1}$ which are determined by the
first $l$ values of the correlation function of an autoregressive
sequence of order $l$ start to become rather complicated as $l$
grows.\

A slightly different approach is suggested in \cite{papastawn20}.\ For
a Markov sequence of order $l < k$, given the initial $l$ values
$\alpha_{0\,:\,l-1}$ with $\alpha_0=1$, structure on
$\bm{\alpha}_{l\,:\,k}$ comes from the recurrence
\begin{align}
\alpha_t = c \bigg\{\sum_{i=1}^{d}\gamma_i(\gamma_i\alpha_{t-i})^{\delta}\bigg\}^{1/\delta}, \quad d \leq t \leq k,  \label{PapaRec}
\end{align}
with $0 < c^{-\delta} < \sum_{i=1}^k \gamma_{i}^{1+\delta}$,
$\delta \in \mathbb{R}$,
$\bm{\gamma}_{1:l} \in S_{l-1} = \{\bm{\gamma}_{1:l} \in [0,1]^l :
\sum_{i=1}^l\gamma_i = 1 \}$.\ For Markov sequences of order 2, the
recurrence in the correlation approach,
$\alpha_t = \theta_1\alpha_{t-1} + \theta_2\alpha_{t-2},$ can be
written in the form (\ref{PapaRec}) with
$\delta=1, \gamma_1 = (\theta_1 -
\sqrt{\theta_1\theta_2})/(\theta_1-\theta_2), \gamma_2 = 1 - \gamma_1$
and
$c = (\theta_1 - \theta_2)^2/(\theta_1 - 2\sqrt{\theta_1\theta_2} +
\theta_2)$.\ There are two points of contrast with the approach using
correlation functions to induce structure on $\bm{\alpha}_{1\,:\,k}$.\
Firstly, now the initial conditions $\bm{\alpha}_{0\,:\,l-1}$ are
regarded as free parameters in $[-1, 1]$ rather than being
parameterized in terms of the parameters of an autoregressive model.\
This reduction in complexity is compensated for by the more complex
recurrence (\ref{PapaRec}).\ In the case of an order 2 Markov sequence
we have five parameters $(\alpha_1, \beta, c, \delta, \gamma_1)$ in
contrast to three parameters when using the correlation function
approach.\ In an implementation of this model, we also need to respect
the parameter constraint $\bm{\gamma}_{1\,:\,l} \in S_{l-1}$.\ In the
case where $l=2$ this is trivial to achieve: we simply confine
$\gamma_1$ to the interval $(0,1),$ e.g., via a logit transform, and
then set $\gamma_2 = 1 - \gamma_1$.\ More generally, we may consider
reparameterizing in terms of $\bm{\Gamma}_{1\,:\,l} \in \mathbb{R}^l$
where
\begin{equation}
\gamma_i = \exp(\Gamma_i)\Big/\sum_{i=1}^l\exp(\Gamma_i), \quad 1\leq i \leq l,
\end{equation}
under which we clearly have $\bm{\gamma}_{1\,:\,l} \in S_{l-1}$.\
However, $\bm{\Gamma}_{1:l}$ is not identifiable as
$\bm{\Gamma}_{1:l}$ and $\bm{\Gamma}_{1:l} + x$ give rise to the same
$\bm{\gamma}_{1\,:\,l}$ for any $x\in \mathbb{R}$.\ This may be dealt
with by adding a sum to zero identifiability constraint,
$\sum_{i=1}^{l}\Gamma_i =0$.\ An identifiable parameterization is then
\begin{equation}
  \gamma_i =
  \begin{cases}
    \dfrac{\exp(\Gamma_i)}{\sum_{i=1}^{l-1}\exp(\Gamma_i)  + \exp(-\sum_{i=1}^{l-1}\Gamma_i)}, \quad 1 \leq i \leq l-1, \\
    \\
    \dfrac{\exp(-\sum_{i=1}^{l-1}\Gamma_i)}{\sum_{i=1}^{l-1}\exp(\Gamma_i) + \exp(-\sum_{i=1}^{l-1}\Gamma_i)}, \quad i=l, 
  \end{cases}
\end{equation}
where $\bm{\Gamma}_{1\,:\,l-1} \in \mathbb{R}^{l-1}$.

\subsection{Examples: semi-parametric approach}  \label{ExSemiParam}

In this section we consider semi-parametric modelling as described in
Section \ref{SemiParamSec} for two order 1 and one order 2 Markov time
series.\ The first example we consider is an autoregressive model with
Gaussian copula.\ Let $Y_0 \sim N(0,1)$ and
\begin{equation}
Y_{n+1} = \rho\,Y_n + \epsilon_n, \quad n \geq 0, \,\, |\rho| < 1,  \label{GaussAR1}
\end{equation}
where $\epsilon_n \sim N(0, 1 - \rho^2$) with $\{\epsilon_n\}$ and
$\{Y_n\}$ independent.\ We transform $\{Y_n\}_{n=0}^{\infty}$ on to
Laplace margins via (\ref{MarginTrans}) with $F=\Phi$ the univariate
Gaussian distribution function.\ Here, as in all examples in this
section, we avoid fitting the marginal model (\ref{MargModSemiParam})
by using the known form of $F$ to obtain exactly Laplace marginals.\
Hence in the simulations that follow we neglect the effect of any
uncertainty from the marginal model.

In our second example, $(X_n, X_{n+1})$ have an inverted logistic
copula.\ Let $Y_0$ be a unit exponential random variable and let the
joint survival function of $(Y_n, Y_{n+1})$ be
\begin{equation}
\bar{F}(y_n, y_{n+1}) = \exp\{-(y_n^{1/\gamma} + y_{n+1}^{1/\gamma})^{\gamma}\}, \quad y_n, y_{n+1} \geq 0,   \label{InvLogistic}
\end{equation}
for $\gamma \in (0, 1]$ and $n \geq 0$.\ We map
$\{Y_n\}_{n=0}^{\infty}$ on to Laplace margins via (\ref{MarginTrans})
where $F(y) = 1 - e^{-y}, y>0$, is the unit exponential distribution
function.

\begin{figure}[htbp!]
  \centering
  \includegraphics[scale=0.45]{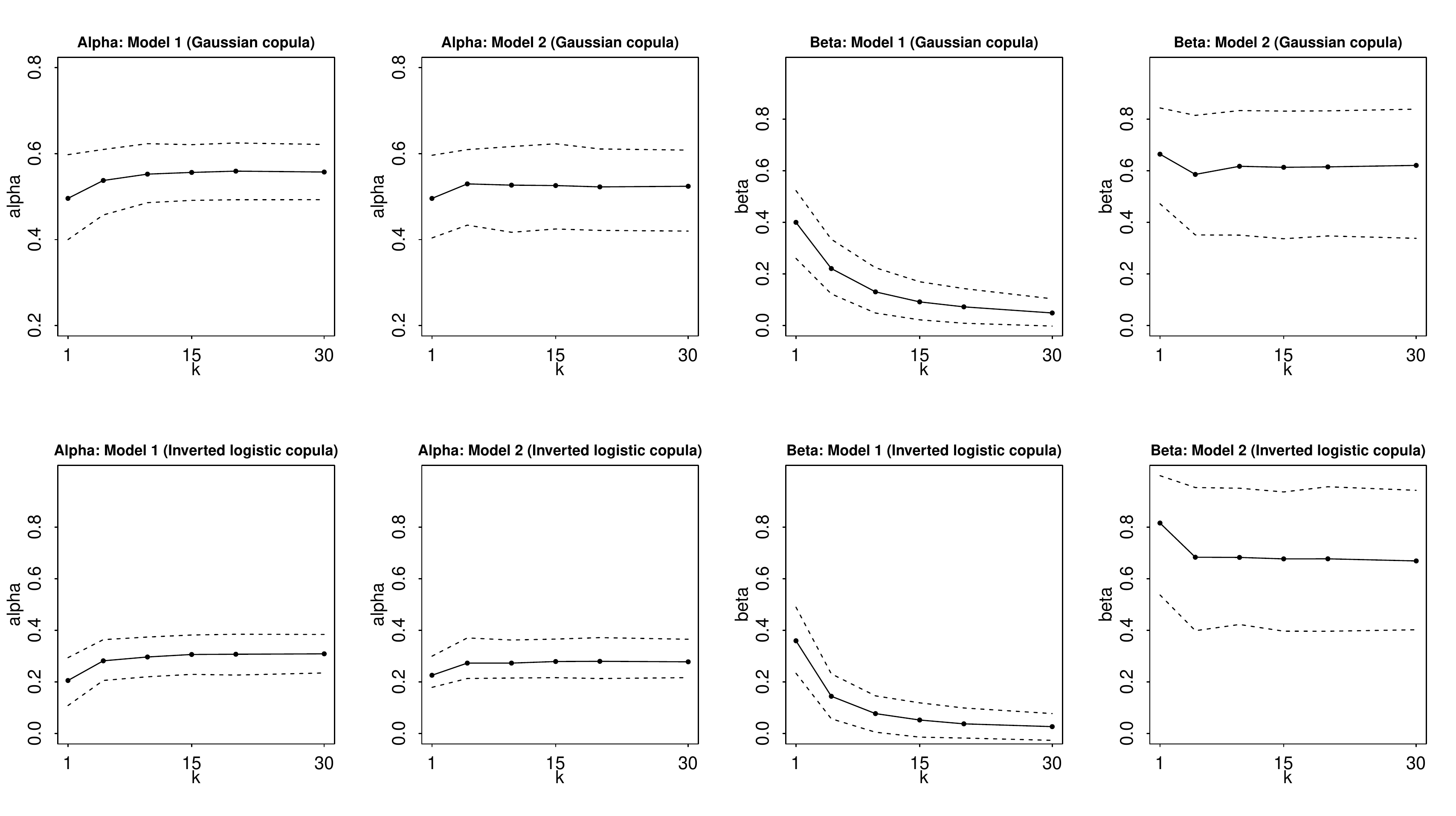}  
  \caption{Plots showing how the estimates $\alpha$ and $\beta$ vary
    with the value of the block length $k$ and different normings for
    Gaussian copula AR(1) (\ref{GaussAR1}) (top row) and inverted
    logistic copula (\ref{InvLogistic}) (bottom row) based on 1000
    realizations of each process.\ The solid lines connect the median
    estimates over all 1000 realizations for different block lengths
    $k$, whereas the broken lines show the 0.025 and 0.975 empirical
    quantiles of the estimates.}
\label{Stability1}
\end{figure}

\begin{figure}[htbp!]
  \centering
  \includegraphics[width=17.8cm]{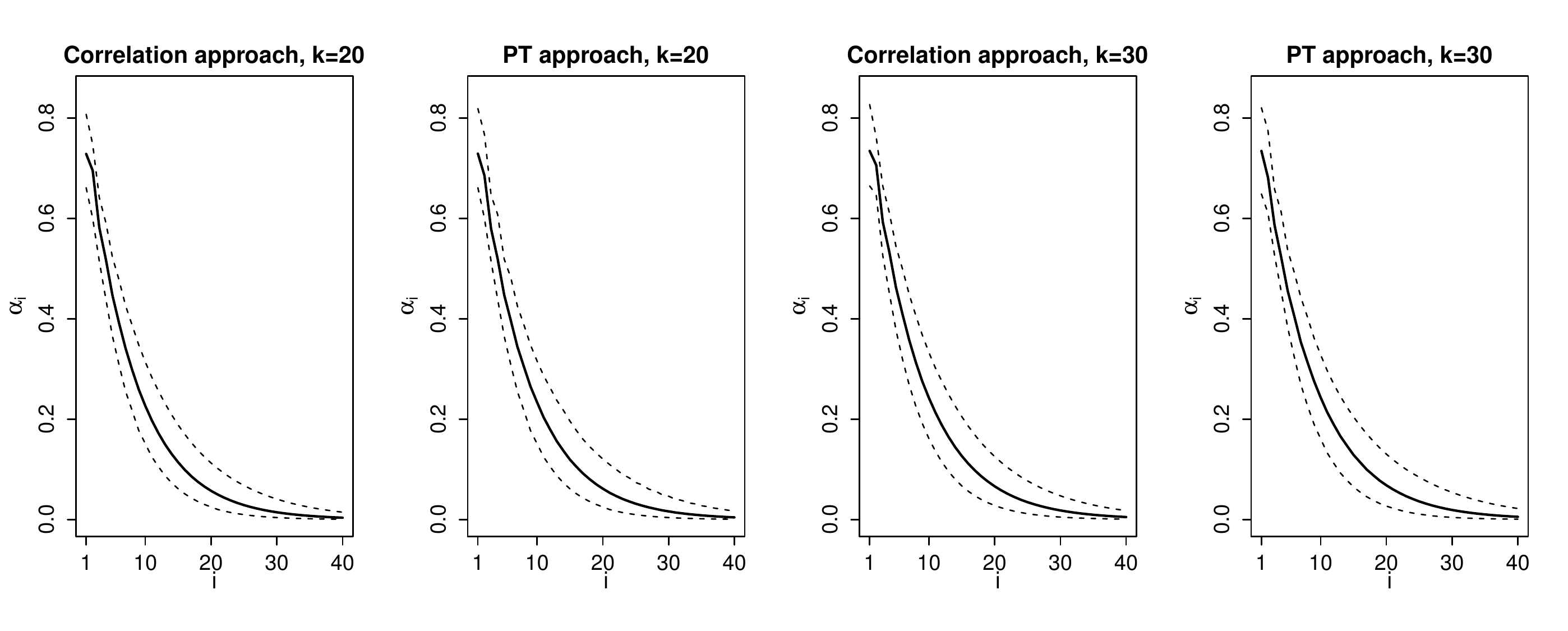}  
  \caption{Plots showing how the estimates of $\alpha_i$ vary with lag
    $i$ for the Markov order 2 Gaussian copula time series
    (\ref{GaussAR2}), using the correlation function approach with
    recurrence (\ref{Cor2}) and the Papastathopoulos-Tawn (PT)
    approach with recurrence (\ref{PapaRec}) based on 500 realizations
    of the process.\ The solid lines connect the median estimates over
    all 500 realizations whereas the broken lines show the 0.025 and
    0.975 empirical quantiles of the estimates.}
\label{AllAlphas_AR2}
\end{figure}

\begin{figure}
  \centering
  \includegraphics[scale=0.5]{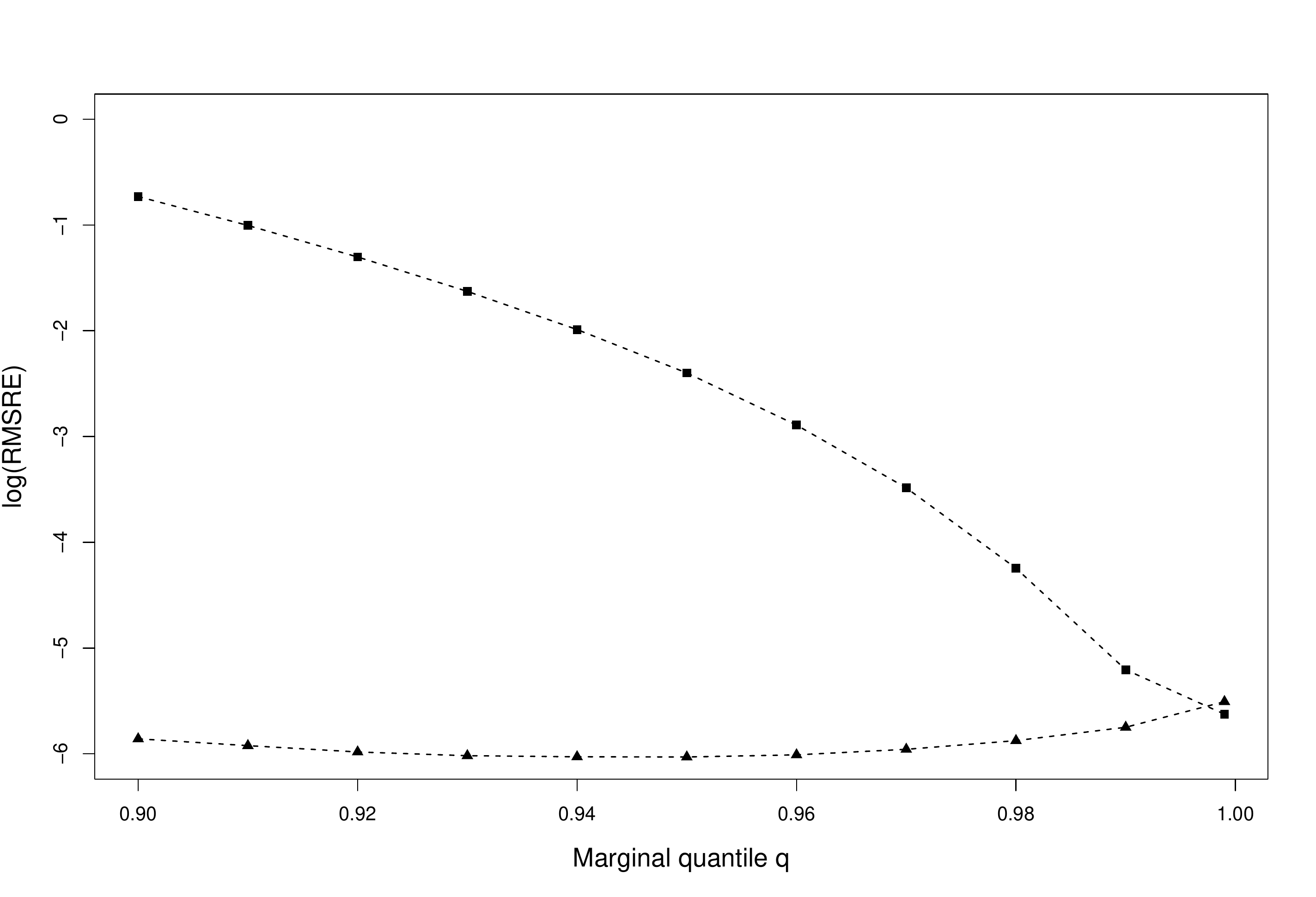}  
  \caption{Plot showing how the log of the root mean squared relative
    error (RMSRE) of estimates of $\theta(v,20$) vary with threshold
    $v = F^{-1}(q)$ for one step method (squares) and $k$-steps method
    (triangles), using $k=19$ and Model 2 normings, in the Gaussian
    copula time series model.\ Estimates are based on $2\times 10^4$
    realizations of the process.}
\label{theta_k20}
\end{figure}

Although both processes are asymptotically independent, the process
with inverted logistic copula requires only a scale normalization,
i.e., the true value of $\alpha$ for this process equals zero whereas
for the Gaussian copula $\alpha = \rho^2$.\ We consider the particular
cases of these two processes when $\rho = 0.7$ and $\alpha = 0.5$ and
investigate how the value of $k$ used in Models 1 and 2 may influence
the estimated values of $\alpha$ and $\beta$.\ We performed a Monte
Carlo study based on 1000 realizations of each process of length
$10^5$ and took our threshold $u$ for identifying exceedances to be
the 0.95 quantile of a standard Laplace distribution.\ For each
realization, we estimated the values of $\alpha$ and $\beta$ using the
semi-parametric approach described in Section \ref{SemiParamSec} for
$k = 1, 5, 10, 15, 20, 30$.\ The median estimates together with the
0.025 and 0.975 empirical quantiles of the estimates are shown for
both copulas and Model 1 and 2 normings in Figure \ref{Stability1}.\
For both copulas and choice of normings, the estimates of $\alpha$ are
stable under different choices for the block length $k$ and vary very
little for $k > 1$.\ We see that under the Model 1 normings there is
noticeable bias introduced in the estimation of $\beta$ as $k$
increases, in particular the estimates seem to be converging to zero.\
The limiting theoretical values of $\beta$ under Model 1 normings are
0.5 for both Gaussian and inverted logistic copulas.\ This increasing
bias in the estimates of $\beta$ at larger values of $k$ is
compensated for by a larger scale in the estimated residual vector
$\bm Z_{1\,:\,k}$ so that reasonable estimates of quantities of
interest using Model 1 may still be obtained.\ The estimates of
$\beta$ under Model 2 normings are much more stable but at the cost of
increasing sampling variability, although the variability does not
show strong dependence on $k$.\ Similar results are found when varying
the sample size and threshold $u$ used to identify exceedances.
To compare the two approaches, (\ref{Cor2}) and (\ref{PapaRec}), to
inducing structure on $\bm{\alpha}_{1\,:\,k}$, for higher-order Markov
sequences we simulated 500 realizations of length $10^5$ of the
Gaussian autoregressive order 2 model
\begin{align}
& Y_n = 0.6Y_{n-1} + 0.3Y_{n-2} + w_n,  \quad w_n \sim N(0, \sigma^2_w), \,\, n \geq 2,    \label{GaussAR2}  \\
& (Y_0, Y_1) \sim \text{MVN}(\bm{0}, \Sigma), \quad \Sigma = \begin{pmatrix} 1 & 0.6 \\ 0.6 & 1\end{pmatrix},   \notag
\end{align}
with $\{w_n\}_{n=2}^{\infty}$ independent of
$\{Y_n\}_{n=0}^{\infty}$.\ We take
$\sigma^2_w = 1 - 0.6^2 - 0.3^2 -2(0.6^2\times\,0.3)/(1-0.3) \approx
0.24$ to ensure $Y_n$ is standard normal for $n\geq 0$.\ Each of the
realizations is transformed on to Laplace margins.\ We only simulated
500 realizations as opposed to 1000, as was done for the first-order
Markov sequences due to the increased time it takes to fit the model
of \cite{papastawn20} in (\ref{PapaRec}).\ Instead of presenting
results for all parameters in both models, we consider just the fitted
values of $\bm{\alpha}_{1\,:\,k}$ which are functions of all model
parameters, except $\beta$, via (\ref{Cor2}) and (\ref{PapaRec}).\ The
median, 0.025 and 0.975 empirical quantiles are shown in Figure
\ref{AllAlphas_AR2}.\ The experiment was repeated using block lengths
of $k$ equal to 20 and 30.\ For values of $i$ larger than $k$, the
values of $\hat{\alpha}_i$ were obtained by extrapolation using the
recurrences (\ref{Cor2}) and (\ref{PapaRec}).\ We used only the Model
1 normings as in (\ref{Mod1Norm}).\ The fitted curves are virtually
identical for both methods, and when plotted on the same diagram both
median estimates and quantiles can hardly be distinguished.\ Very
similar estimates for $\beta$ were also obtained from each model.\
When using a block length of $k=20$, the 0.025 and 0.975 quantiles of
estimates of $\beta$ obtained by using the recurrences (\ref{Cor2})
and (\ref{PapaRec}) were $[0.158, 0.379]$ and $[0.157, 0.390]$
respectively.\ The corresponding intervals when $k=30$ were
$[0.095, 0.323]$ and $[0.102, 0.308]$ for (\ref{Cor2}) and
(\ref{PapaRec}) respectively.\ The main difference between the two
methods is that the approach using the recurrence (\ref{PapaRec}),
with its two extra parameters, takes considerably longer to fit.\ For
fitting Markov models with order more than 2, it may be useful to
replace the working assumption of $\delta$-Laplace margins for the
residual vector with Gaussian margins in order to speed up the fitting
procedure.
To illustrate the possible utility of simulating a large block forward
from an extreme event, we performed an experiment to estimate the
subasymptotic extremal index $\theta(v,d)$ defined in (\ref{SAEI})
where $d=20$ for a range of thresholds $v$ in the Gaussian copula
AR(1) model (\ref{GaussAR1}).\ The thresholds considered correspond to
the 0.9 up to 0.99 marginal Laplace quantiles in increments of 0.01
together with the 0.999 quantile.\ As we need to be able to simulate
19 steps forward from the first exceedance, we used a value of
$k = 19$ and estimated $\alpha$ and $\beta$ using the semi-parametric
method of Section (\ref{SemiParamSec}) and Model 2 normings.\ We
compare this with the case where $k=1$ and a working Gaussian
assumption on the residual $Z_{t+1}$ in (\ref{WT_rec}) was used which
then corresponds to the method used in \cite{winttawn16}.\ For this
latter method, Algorithm \ref{SAEI_Alg} cannot be used to estimate
$\theta(v,20)$ and instead Algorithm 1 of \cite{winttawn16} is used.\
This involves simulating forward from the initial exceedance one step
at a time using the recurrence (\ref{WT_rec}).\ We simulated
$2\times 10^4$ sequences of length $2\times 10^4$ and used a value of
$n=5\times 10^4$ in Algorithm \ref{SAEI_Alg} for the case $k=19$ and
similarly for when $k=1$.\ We used a threshold of $u$ equal to the 0.9
marginal Laplace quantile to identify exceedances.\ For each sequence
we use the fitted values of $\alpha$ and $\beta$ to estimate
$\theta(v,20)$ and measure the quality of the estimate via the squared
relative error.\ The overall performance of each of the two methods
for a given threshold is then taken to be the root mean squared
relative error (RMSRE).\ That is, if for a given threshold $v$, a
given method produces estimates $p_1,\ldots, p_M$, where
$M=2\times 10^4$, we report the value
$\big[M^{-1}\sum_{i=1}^M(p_i/p - 1)^2\big]^{1/2}$ where
$p=\theta(v,20)$.\ Figure \ref{theta_k20} shows how the log of RMSRE
varies with threshold $v$ for both methods.\ For all thresholds up to
the 0.98 marginal quantile, the $k$-steps method where we simulate
blocks jointly of length 19 forward from the first exceedance
dominates the method where we simulate forward one step at a time.\
The $k$-steps method displays extremely stable performance at all
thresholds with very little discernible differences in the RMSRE
values.\ The difference between the two methods however diminishes as
we increase the threshold $v$ and by the time we reach the 0.99
quantile, the two methods are virtually indistinguishable.\ This may
simply be due to the asymptotic independence of the process, which
ensures that $\theta(v,20) \to 1$ as $v\to \infty,$ and so at large
thresholds $v$, $p_i / p \approx 1$ and RMSE $\approx 0$ for both
methods.\ A further possible factor explaining the relatively poor
performance of the Winter--Tawn model at lower thresholds is due to a
detail of the simulation scheme used for this model.\ For this model,
we simulate forward from an initial threshold exceedance one step at a
time using the recurrence (\ref{WT_rec}).\ If $X_t < 0$ for some $t$
then $X_{t+1}$ will be undefined according to (\ref{WT_rec}) as
$\beta < 1$.\ In this case, Algorithm 1 of \cite{winttawn16} sets all
subsequent values within the cluster to zero.\ The idea here is that
as zero corresponds to the median value on the Laplace scale, if we
drop below this value at some point then we do not expect subsequent
values to be extreme and so setting these to zero should have little
effect on estimated cluster functionals.\ However, at moderately high
thresholds such as the 0.9 quantile, even after dropping below this
threshold, there is a small but non-negligible probability of again
exceeding the threshold.\ Thus setting all subsequent values to the
marginal median may be leading to significant bias in the estimates.

\subsection{Examples: parametric approach}   \label{ExParam}

For each of the three time series models, (\ref{GaussAR1}),
(\ref{InvLogistic}) and (\ref{GaussAR2}), from Section
\ref{ExSemiParam}, we now consider the possibility of identifying
parametric forms for, $\mu_i, \sigma_i$ and $\delta_i$, the
$\delta$-Laplace parameters associated to the lag $i$ residual term
$Z_{t+i\mid t}$.

\begin{figure}[htbp!]
  \centering
  \includegraphics[scale=0.54]{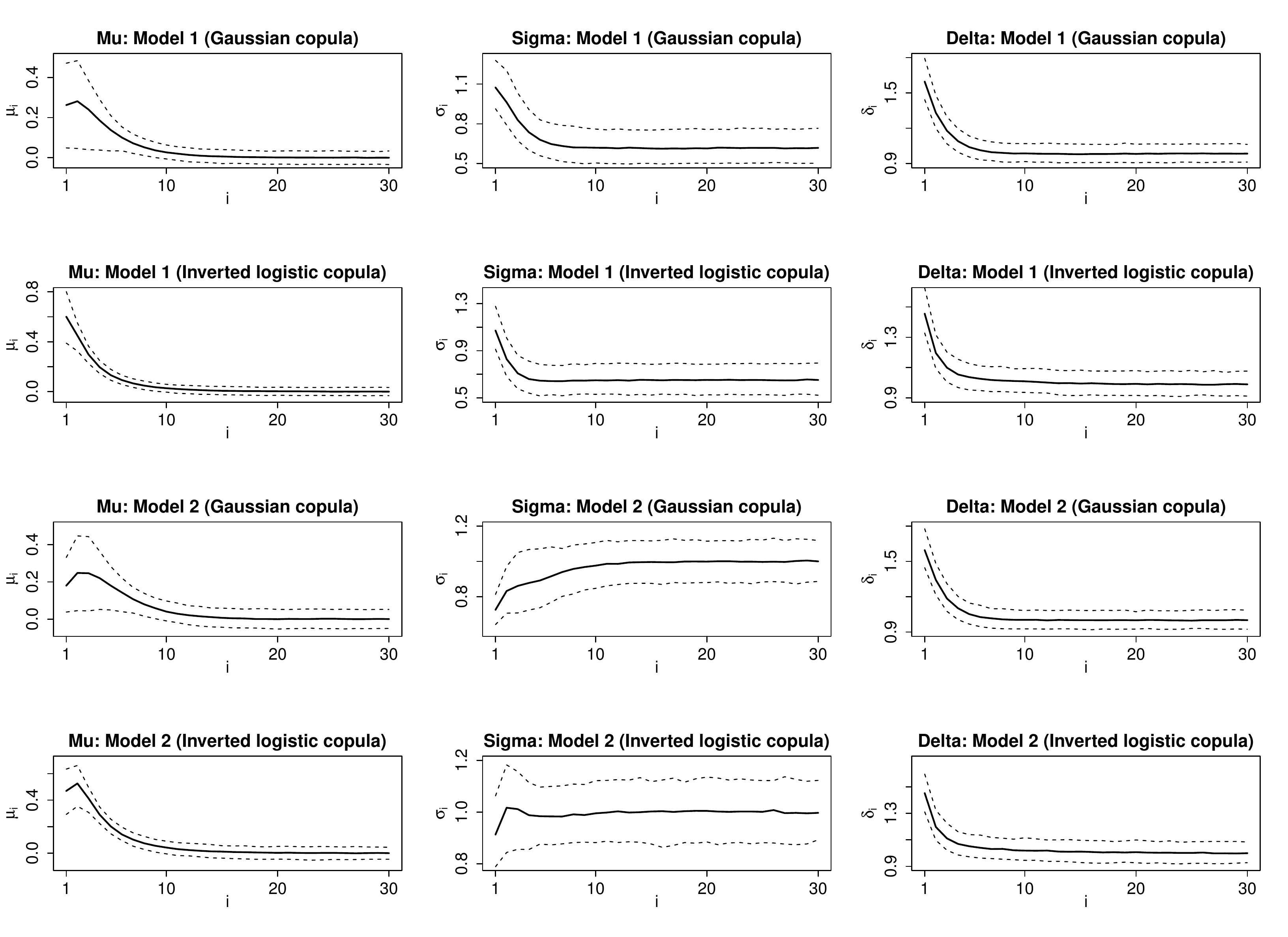}  
  \caption{Plots showing how the $\delta$-Laplace parameters
    $\mu_i, \sigma_i$ and $\delta_i$ vary with $i$ under Model 1 (top
    two rows) and Model 2 (bottom two rows) normings for Gaussian and
    inverted logistic copula models based on 1000 realizations of each
    process.\ The solid lines connect the median estimates over all
    1000 realizations whereas the broken lines show the 0.025 and
    0.975 empirical quantiles of the estimates.}
\label{MuSigDel_Mod1}
\end{figure}

Focusing, for now, on the first-order models (\ref{GaussAR1}) and
(\ref{InvLogistic}), we took the fitted values of $\alpha$ and
$\beta$, say $\hat{\alpha}$ and $\hat{\beta},$ from our Monte Carlo
experiment described in Section \ref{ExSemiParam} and calculated the
empirical lag $i$ residuals as
\begin{equation}
\hat{Z}_i = \frac{X_{t+i\mid t} - \hat{\alpha}^iX_{t}}{X_t^{\hat{\beta}}}, \quad t \in T_u  \label{Res1}
\end{equation}
for Model 1 and 
\begin{equation}
\hat{Z}_i = \frac{X_{t+i\mid t} - \hat{\alpha}^iX_{t}}{1 + (\hat{\alpha}^iX_t)^{\hat{\beta}} }, \quad t \in T_u    \label{Res2}
\end{equation}
for Model 2 normings.\ This was repeated separately for each value of
$k = 1, 5, 10, 15, 20, 30$.\ From these residuals we then estimated
the $\delta$-Laplace parameters, $\mu_i, \sigma_i$ and $\delta_i$, of
the residual $Z_{t+i\mid t}$, via maximum likelihood.\ This was
repeated for lags $i$ from 1 up to 30 inclusive.\ Figure
\ref{MuSigDel_Mod1} shows how the fitted $\delta$-Laplace parameters
of $Z_{t+i\mid t}$ vary with lag $i$ using Model 1 and Model 2
normings for both Gaussian and inverted logistic copulas in the case
where we used $k=1$.\ We see that the same parametric forms for
$\mu_i, \sigma_i$ and $\delta_i$ could be used for both copulas
although the choice of norming appears to have some effect, most
noticeably on the scale parameter $\sigma_i$.

Focusing, for now, on Model 1 normings, exponential decaying models
for each of the $\delta$-Laplace parameters appear an appropriate way
to describe how the processes return to the body after their
excursions in the tail.\ Although there is a hint of a turning point
near the start of the $\mu_i$ curve for the Gaussian copula it is
sufficiently small that it could possibly be ignored.\ The values of
$\mu_i$ and $\delta_i$ for large values of $i$ stabilize to 0 and 1
respectively as expected; these are the values of the marginal Laplace
distribution.\ The value of $\sigma_i$ for both copulas appears to
converge to a value between 0.6 and 0.7.\ To try and understand the
behaviour of $\sigma_i$ at large lags $i$, we recall the model based
assumption
$X_{t+i\mid t}~|~X_t > u\sim \delta\text{Laplace}(\alpha^iX_t +
X_t^{\beta}\mu_i, X_t^{\beta}\sigma_i, \delta_i)$.\ Thus, we see that
provided $|\alpha | < 1$, the location and shape parameters of the
distribution of $X_{t+i\mid t}~|~X_t > u,$ for large $i$ will not be
approximately those of a standard Laplace distribution unless $\mu_i$
and $\delta_i$ are such that $\mu_i \approx 0$ and
$\delta_i \approx 1 $ for large $i$.\ However, for the scale parameter
of the distribution of $X_{t+i\mid t}~|~X_t > u$ to match that of a
standard Laplace distribution we would require
$\sigma_i \approx X_t^{-\beta}$ for large $i$ and this clearly cannot
hold for all $t \in T_u$.\ One possible workaround here is to specify
a parametric form for $\sigma_i$ so that for large $i$,
$\sigma_i \approx \mathbb{E}(X_t \mid X_t > u)^{-\beta} = (1 +
u)^{-\beta}$ due to the exponential right tail of Laplace
distribution.\ Thus a possible parameterization under the Model 1
normings for both copulas is
\begin{align}
\mu_{i+1} &= Ae^{-Bi}, \notag \\
\sigma_{i+1} & = (1 + u)^{-\beta}(1 + Ce^{-Di}),  \label{Parameterization1} \\
\delta_{i+1} &= 1 + Ee^{-Fi},  \notag
\end{align}
for $A, B, C, D, E, F > 0$ and $i\geq 0$.\ When $u$ is the 0.95
quantile of a standard Laplace distribution and $\beta = 0.35$
(approximately the median estimate of $\beta$ when using $k=1$ for
both copulas) then $(1 + u)^{-\beta} \approx 0.66$, which explains the
convergence of $\sigma_i$ to approximately this value in Figure
\ref{MuSigDel_Mod1}.\ Similar curves (not shown) for $\mu_i, \sigma_i$
and $\delta_i$ appear when repeating this experiment with larger $k$.\
The main difference is that the estimates of $\beta$ decrease to zero
for large $k$ and so the factor $(1+u)^{-\beta}$ in
(\ref{Parameterization1}) is approximately 1.\ Consequently, we find
$\sigma_i$ converges to approximately 1.

For Model 2 normings, we see $\mu_i \to 0, \sigma_i \to 1$ and
$\delta_i\to 1$, and so we don't need the factor $(1+u)^{-\beta}$ as
in (\ref{Parameterization1}).\ This is due to the norming used for
Model 2 which ensures the $\delta$-Laplace scale parameter of
$X_{t+i}~|~X_t>u$ is $\{1 + (\alpha^iX_t)^{\beta} \}\sigma_i$ and
provided $\alpha \in (0, 1)$, the term $(\alpha^iX_t)^{\beta}$ goes to
zero with probability 1.\ Another notable difference when using the
Model 2 normings is that the behaviour of $\sigma_i$ appears to have
been inverted in comparison to Model 1.\ It now increases
monotonically before stabilizing to one.\ A possible parameterization
under the Model 2 normings for both copulas is
\begin{align}
\mu_{i+1} &= Ae^{-Bi}, \notag \\
\sigma_{i+1} & = 1 + Ce^{-Di},  \label{Parameterization2} \\
\delta_{i+1} &= 1 + Ee^{-Fi}  \notag
\end{align}
for $i\geq 0$, where now $C < 0$.\ There appears to be a slightly more
pronounced mode near the beginning of the $\mu_i$ curves.\ If we
wished to accommodate this feature, this could be accomplished by
including an additional parameter, e.g.,
$\mu_{i+1} = Ai^{B-1}e^{-Ci},$ which is proportional to the density
function of a gamma random variable.
We consider fitting the parametric model (\ref{Parameterization1}) for
the Gaussian copula autoregression using the method of Section
\ref{ParamSec} for all 1000 realizations of the process that were used
in the Monte Carlo experiment for semi-parametric estimation.\ This
requires us to specify the correlation matrix $P$ of the Gaussian
copula of the residual vectors $\bm Z_{t+1\,:\,t+k\mid t}, t\in T_u$.\
As discussed in Section \ref{ParamSec} we consider the correlation
matrix of the $(k+1)\mhyphen \textnormal{dimensional}$ random vector
$\bm Z_{0\,:\,k}$ conditioned on $Z_0$.\ As we are working with a
first order Markov sequence, we take $P$ to be the conditional
correlation matrix of a first-order stationary autoregressive model
with autocorrelation parameter $\rho$.\ It is more convenient to give
a specification directly in terms of $Q = P^{-1}$ which is a sparse
(banded) matrix and appears in (\ref{LogCondDens}).\ Moreover,
updating precision matrices upon conditioning is much more
straightforward than for correlation matrices \citep[Theorem
2.5]{rue05}.\ We first construct the $(k+1)\times\,(k+1)$ dimensional
matrix $\widetilde{Q}$ as
\begin{equation}
\widetilde{Q}_{i,j} = 
\begin{cases}
1,   &  \quad \text{if\,\,} (i,j)=(1,1)\,\,\text{or\,\,} (i,j)=(k+1,k+1), \\
1+\rho^2, & \quad \text{if\,\,} i=j, \,\, \text{and\,\,} 2\leq i \leq k, \\
-\rho,  & \quad \text{if\,\,} |i-j| = 1, \\
0, & \quad \text{otherwise}.
\end{cases}
\end{equation}
The matrix $Q$ used in (\ref{LogCondDens}) is then obtained by
deleting the first row and column of $\widetilde{Q}$.\ Although we
were able to maximize the composite likelihood jointly for all
parameters $(\alpha, \beta, A,\ldots,F,\rho)$, it was sufficiently
slow to make a simulation study infeasible.\ For this purpose, we
estimate the parameter vector in two stages as advocated in
\cite{joe97}.\ First we estimate $(\alpha, \beta, A,\ldots,F)$ under
the working assumption of an independence copula on the residual
vector, and then maximize the composite likelihood for $\rho$ assuming
the parameters $(\alpha, \beta, A,\ldots,F)$ are fixed at their
estimated values from the first step.\ This greatly speeds up the time
to fit the model compared to estimating all parameters at once.\ Based
on Figure \ref{MuSigDel_Mod1}, we constrain $A$ and $E$ to the
interval $(0,1)$, $C$ to $(0,3)$ and $B, D $ and $F$ to $(0,5)$ using
scaled logit transformations.\ The 0.025 and 0.975 empirical
quantiles, to two decimal places, of the parameters estimates from the
1000 realizations are shown in Table \ref{ParamQuants}.\ We performed
our estimation for values of the block length $k = $10, 20 and 30.\
The estimates do not show strong dependence of $k$ beyond what was
already noted for $\beta$ as when using the semi-parametric approach
with Model 1 normings.\ The estimates for $B$ in particular are highly
uncertain and the 0.025 and 0.975 quantiles encompass most of the
interval $(0,5)$.\ A value of $B=5$ would correspond to extremely
rapid decay of $\mu_i$ with $i$ which does not seem to be supported by
the plots in Figure \ref{MuSigDel_Mod1}.\ One possibility for the high
uncertainty in the estimates of $B$ could be due to a poorly specified
parametric form for $\mu_i$ since as we already noted, there is
evidence of a turning point in Figure \ref{MuSigDel_Mod1} which we
have ignored.\ \cite{shooter19} also report difficulties with
likelihood based estimation for parametric $\delta$-Laplace models and
find better results using Bayesian methods.\ A similar approach in the
time series setting may be useful in future work.\ Estimates for the
other parameters seem reasonable and the copula correlation parameter
$\rho$ in particular is estimated with precision and is very stable as
we vary the block length $k$.

\begin{figure}[htbp!]
  \centering
  \includegraphics[scale=0.5]{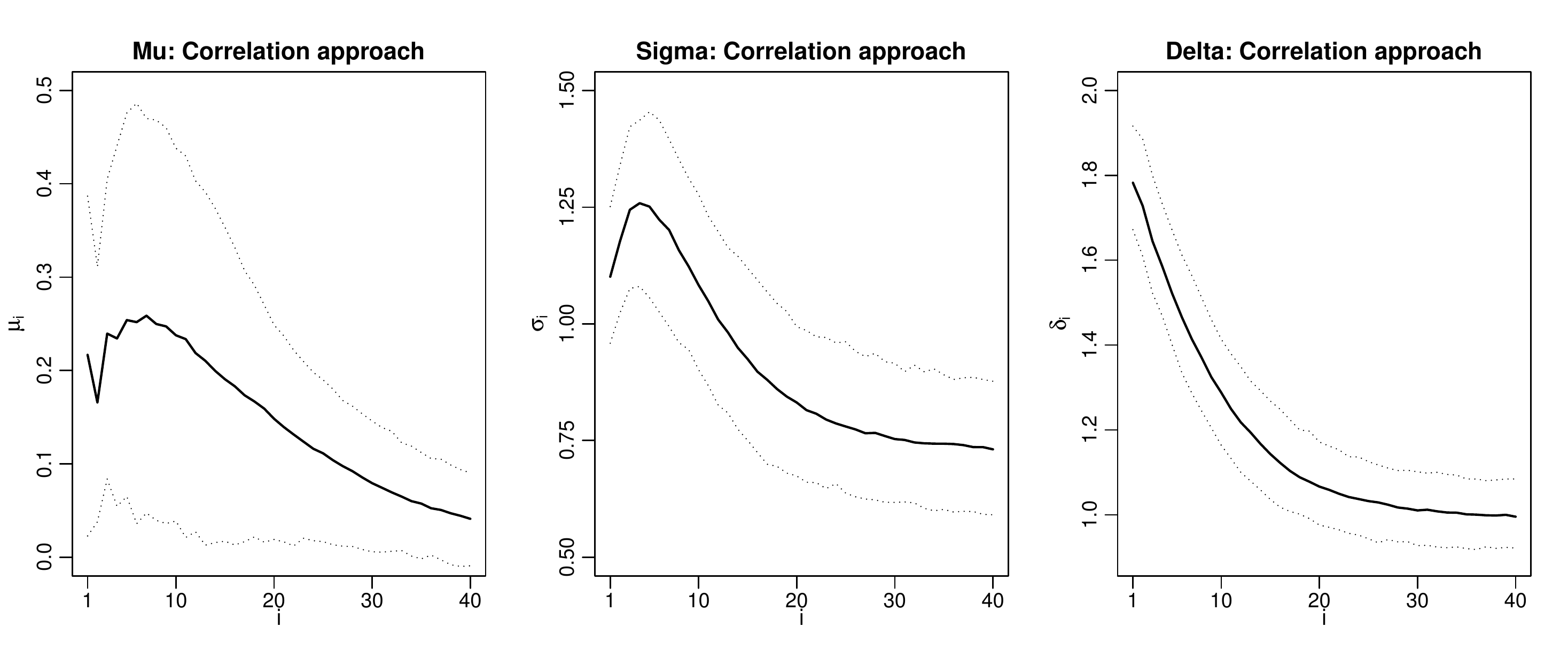}  
  \caption{Plots showing how estimates of the $\delta$-Laplace
    parameters $\mu_i, \sigma_i$ and $\delta_i$ of the residual vector
    vary with lag $i$ for the Markov order 2 Gaussian copula time
    series (\ref{GaussAR2}), using the correlation function approach
    with recurrence (\ref{Cor2}) and block length $k=20$ based on 500
    realizations of the process.\ The solid lines connect the median
    estimates over all 500 realizations whereas the broken lines show
    the 0.025 and 0.975 empirical quantiles of the estimates.\
    Virtually identical curves (not shown) were obtained when using
    Papastathopoulos-Tawn model.}
\label{MuSigDelta_AR2}
\end{figure}

\begin{table} 
\centering
\caption{0.025 and 0.975 quantiles, to two decimal places, of the
  sampling distributions for parameters in the parametric model
  (\ref{Parameterization1}) and Gaussian copula autoregressive model.\
  These are based on 1000 realizations of the process for different
  block lengths $k$.}
 \label{T1}
{\begin{tabular}{|l  | l  l  l  l  l  l  l  l  l | }
 \hline 
  $k$ &  $\alpha$ & $\beta$ &  $A$  & $B$ & $C$ & $D$ & $E$ & $F$ & $\rho$    \\
\hline\hline
 $10$ & $.54,.63$ & $.08, .20$ & $ .00,.35 $ & $ .10,5.00 $ &  $ .63,1.03$ & $.35, 1.21$   & $.45,.99$ & $.44,2.46$ & $.65,.66$   \\
$20$  & $.54,.64$ & $.03,.12$ &    $.00,.31$ & $.13,5.00$ &   $.65,0.99$      &  $.35, .84$   & $.48,.95$ & $.44,1.27$ & $.67,.68$  \\
$30$ &  $.54,.63$ & $.01,. 09$  &   $.00, .32$ & $.15,4.99$ &   $.63,1.03$   &  $.35,1.18$   & $.46,.99$  & $.42,2.08$ & $.68,.69$  \\   
\hline
 \end{tabular}   \label{ParamQuants}}
\end{table}
To test whether parametric models for the $\delta$-Laplace parameters
of the residual vectors may be possible for the second order Markov
time series (\ref{GaussAR2}), we carried out the same procedure as
described for the first-order Markov sequences.\ That is, from a given
model fit from the Monte Carlo experiment of Section
\ref{ExSemiParam}, we calculated the lag $i$ residuals as in
(\ref{Res1}) and estimated the $\delta$-Laplace parameters
$\mu_i, \sigma_i$ and $\delta_i$ by maximum likelihood.\ Figure
\ref{MuSigDelta_AR2} shows the median estimate with the 0.025 and
0.975 empirical quantiles of the estimates when using the correlation
function approach with recurrence (\ref{Cor2}) and block length
$k=20$.\ The approach of \cite{papastawn20} using recurrence
(\ref{PapaRec}) gave essentially the same curves as did using a block
length of $k=30$.\ The curves for $\mu_i$ and $\sigma_i$ look more
complex than in the order 1 case and have a clear mode that would need
to be modelled although the same model for $\delta_i$ as in the first
order case seems appropriate.

\section{Data application}  \label{DataApp}

\subsection{Data} 

In this section we illustrate our methodology on a time series of
daily maximum temperature measurements from Orleans, France, during
the years 1946-2012.\ This is the same data set that is analysed in
\cite{winttawn16} and \cite{Winter2017}.\ We work only with the summer
months June-August and assume that these months from separate years
constitute approximately independent realizations of a stationary
process.\ There are four missing values which all occur in different
years.\ We imputed missing values with a Kalman filter using the R
\citep{R21} package \texttt{imputeTS} \citep{imputeTS}.

\subsection{Methods and model}  \label{MethModSec}

\cite{winttawn16} fit a first-order Markov model to the Orleans data
based on the fact that the partial autocorrelation function is not
significantly different from zero at lag 1.\ Based on more refined
diagnostics, similar to those described in Section \ref{DiagSec},
\cite{Winter2017} fit a Markov model of order 3.\ Both papers find
asymptotically independent models to be appropriate for the data.\
Here, we consider asymptotically independent models that assume the
structure of the parameters $\bm{\alpha}_{1\,:\,k}$ is that of an
autocorrelation function of a stationary autoregressive sequence as
described in Section \ref{Mark1}.\ We fit autocorrelation functions of
order 1,2 and 3 using the semi-parametric approach and Model 1
normings.\ We also fit the first-order Markov model of
\cite{winttawn16} for the sake of comparing it's predictions with our
own models.

We will consider simulation and estimation of several functions of
daily maximum temperature in Orleans over a three week period
conditional on there being an exceedance of 35\textdegree{}C at the
beginning of the period, i.e., we will only consider forward
simulation as described in Section \ref{ForSim}.\ A temperature of
35\textdegree{}C corresponds approximately to the one year return
level, i.e., this temperature is exceeded on average once per year.\
As discussed in \cite{winttawn16}, a period of three consecutive days
with mean daily maximum temperature in excess of 35\textdegree{}C may
lead to excess mortality in Orleans between 17\% and 47\%.

As we focus on a three week period, we need to be able to simulate 20
steps ahead from the initial exceedance, and so we use a block length
of $k=20$ when fitting our model.\ An extra complication arises with
model fitting that was less present in the simulated examples of
Section \ref{Ex}, due to the segmented structure of the data.\ In
particular, if $X_t$ is an exceedance of our fitting threshold $u$, on
the Laplace scale, that occurs after August 12 , then there will be
less than 20 successive values available that year and so we cannot
get a composite likelihood contribution from $\bm{X}_{t+1\,:\,t+20}$.\
In such cases, we still allow a composite likelihood contribution,
through $\bm{X}_{t+1\,:\,t+m}$ where $m < 20$ is the maximum lag
available.

In order to quantify uncertainty in parameter estimates and other
quantities of interest, we use a moving block bootstrap with block
length 20 to simulate 1000 replicate data sets of the same length and
structure as the original data set.\ From these replicate data sets
approximate standard errors may be obtained for any estimates by
calculating the standard deviation of estimates across all bootstrap
samples.\ Similarly, bootstrapped 95\% confidence intervals are
obtained from the 0.025 and 0.975 empirical quantiles of estimates
across all samples.\ In Section \ref{ResultsSec}, due to the high
computational cost of evaluating the cluster functional probabilities,
we use a reduced number, namely 300, bootstrap samples to evaluate
uncertainties in our estimates.

\subsection{Diagnostics} \label{DiagSec2}

\begin{figure}[htbp!]
  \centering \includegraphics[scale=0.53, trim=20 20 20
  20]{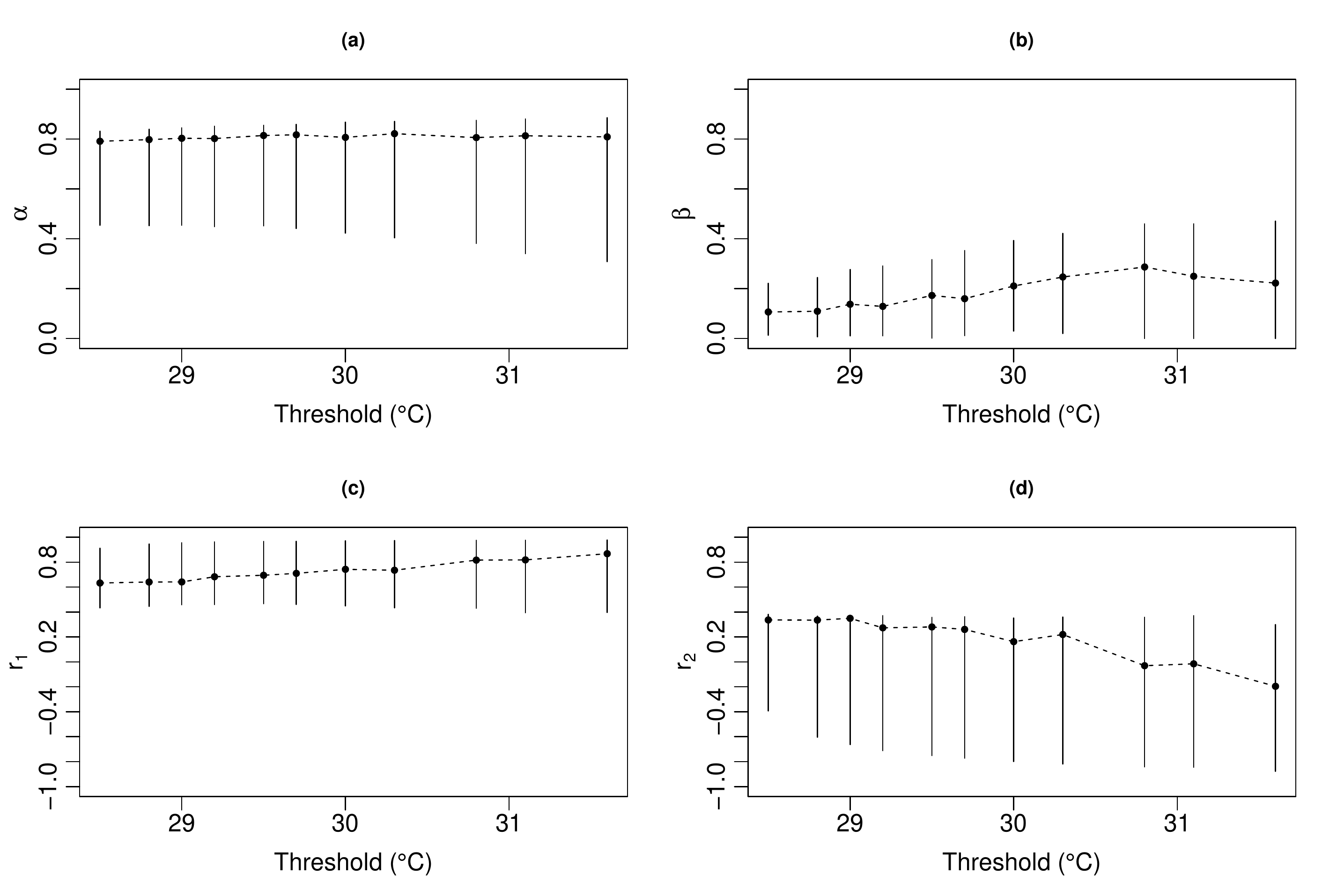}
  \caption{Plots showing parameter estimates and 95\% bootstrap
    confidence intervals for different thresholds used to identify
    exceedances.\ The top row shows estimates of (a) $\alpha$ and (b)
    $\beta$ for the first-order Markov correlation model.\ The second
    row shows estimates of (c) $r_1$ and (d) $r_2$ for the second
    order Markov correlation model.}
\label{ParamStabPlot}
\end{figure}

The marginal model from Section \ref{MargMod} was applied to the
Orleans temperature data using a threshold of $u^* = 29.7$ which
corresponds to the 0.9 marginal empirical quantile.\ The generalized
Pareto distribution was fit to excesses of this threshold to obtain
parameter estimates of $\hat{\sigma} = 2.801\, (0.185)$ and
$\hat{\xi} = -0.192\, (0.042)$, with estimated standard errors in
parentheses.\ Parameter stability plots are given in \cite{winttawn16}
that justify this threshold choice.\ Parameter stability plots for
$\alpha$ and $\beta$ for the first-order correlation model and the
parameters $r_1$ and $r_2$ for the partial autocorrelation
parameterization of the order 2 model are shown in Figure
\ref{ParamStabPlot}.\ The stability plots for $\beta$ were virtually
identical for the correlation models of all orders and the plots for
the parameters of the order 3 model were very similar to the
corresponding parameters of the order 2 model.\ Based on these plots
we select a threshold $u$ on the Laplace scale corresponding to
30\textdegree{}C to fit our conditional models.\ For the first-order
model we obtain estimates of $\hat{\alpha} = 0.807\, (0.124)$ and
$\hat{\beta} = 0.211\, (0.095)$, for the second order model
$\hat{r}_1 = 0.742\, (0.157),$ $\hat{r}_2 = 0.162\, (0.323),$
$\beta = 0.210\, (0.095),$ and for the third order model
$\hat{r}_1=0.764\, (0.137),$ $\hat{r}_2 = -0.048\, (0.286),$
$\hat{r}_3 = 0.163\, (0.338)$ and $\hat{\beta} = 0.209\, (0.095)$.\ We
observe that the approximate 95\% confidence interval for $r_2$ for
the second order model of $[-0.799, 0.352]$ contains zero.\ For the
third order model, the 95\% confidence intervals of $[-0.749, 0.494]$
and $[-0.747, 0.602]$ for $r_2$ and $r_3$ also contain zero.\ Thus we
have some initial evidence that the higher-order structure provided by
the second and third order models may be redundant.\ For the
Winter--Tawn first-order Markov model we obtained estimates of
$\hat{\alpha} = 0.734\,(0.237)$ and $\hat{\beta} = 0.546\,(0.129)$.\
Table \ref{AtLeastTab} (a) compares empirical and model based
estimates of the probability of at least $s$ occurrences of the 0.9
empirical quantile over a three week period given an exceedance of the
threshold at the start of the period, while (b) shows the equivalent
comparison using the 0.95 empirical quantile.\ Table \ref{ChiTab}
compares estimates of the subasymptotic tail dependence measure
$\chi(v, d)$, defined in (\ref{SAE}), with the threshold $v$ set to
the empirical 0.9 and 0.95 empirical quantiles as the lag $d$ varies
from 1 to 10.\ Recall that the rationale for such comparisons is that
at moderately high thresholds such as the 0.9 quantile, empirical
estimates ought to be reasonably accurate, and so we may detect
potential model deficiencies by comparing the estimates they produce
with their empirical counterparts.\ From Table \ref{AtLeastTab} (a),
we see that the correlation function model based estimates of all
orders are very similar to the empirical estimates for each $s$ and in
particular, are always within one standard error of the empirical
estimates.\ Similar comments apply for the higher threshold in Table
\ref{AtLeastTab} (b), The Winter--Tawn model tends to underestimate
the probabilities relative to the empirical estimates and decays too
quickly as $s$ increases.\ Its poor performance is likely in part due
to the problems identified at the end of Section \ref{ExSemiParam} of
simulating from this model at such subasymptotic thresholds and we
would expect such problems to not be so pronounced at much higher
thresholds.\ The correlation model based estimates of $\chi(v, d)$
with $v$ equal to the 0.9 empirical quantile broadly agree with the
empirical estimates; although they tend to overestimate the
probabilities, they are usually within one standard error of the
empirical estimates.\ The same general comment applies in the case of
the 0.95 empirical quantile, while the overestimation is a little more
evident, the correlation model based estimates are usually within two
standard errors of the empirical estimates.\ The Winter--Tawn model
based estimates show strong agreement with the empirical estimates at
short lags $d$ but decay too rapidly.\  At short lags the problem with
the forward simulation scheme for the Winter--Tawn model is not so
evident as when we need to simulate and evaluate cluster functionals
over a much larger block length.

For the probabilities that are estimated in Section \ref{ResultsSec},
Table \ref{AtLeastTab} is the more pertinent diagnostic as it
considers the joint behaviour of all observations over a three week
period whereas estimates of $\chi(v, d)$ only consider the bivariate
behaviour.\ In other applications, the precise form of cluster
functional probabilities to be estimated will dictate which
diagnostics are most relevent.\ Taking all evidence in to account, we
find no compelling evidence to prefer the higher-order correlation
models over the first-order model which is our preferred model.\
However, in Section \ref{ResultsSec} we will present estimates
produced by all models.

An important modelling assumption that we utilize for simulations is
that the residual vector $\bm Z_{t+1\,:\,t+k\mid t}$ is conditionally
independent of $X_t$ given $X_t > u$.\ We assessed the plausibility of
this assumption for the first-order correlation function model via
scatterplots (not shown) of $Z_{t+i\mid t}$, the component of the
residual vector at lag $i$, against $X_t$, $t \in T_u$, for each
$i\in\,1{\,:\,}20$.\ No obvious dependencies were visually apparent.\
Slightly more formally, we calculated the values of Kendall's measure
of association $\tau$ \citep[Section 2.1.9]{joe97} for the pairs
$\{(X_t, Z_{t+i\mid t})\}_{t\in\,T_u}$ for each lag
$i\in\,1{\,:\,}20$.\ All of the estimated values of $\tau$ were of
order $10^{-2}$ suggesting that any dependence between components of
the residual vector and threshold exceedances is very weak.\ We
repeated this procedure for all bootstrap samples.\ The bootstrap
distributions, i.e., histograms, of the fitted values of $\tau$ are
shown in Appendix \ref{AppendixDiag}.\ We obtained 95\% bootstrap
confidence intervals, correcting for multiple comparisons using the
Bonferroni correction.\ All 20 of these intervals contained zero
implying no significant association at the $5\%$ level of
significance.\ Although we have not directly tested for independence
of $\bm Z_{t+1\,:\,t+k\mid t}$ and $X_t$, $t\in T_u$, these informal
diagnostics suggest that such an assumption is reasonable.
\begin{table} [htp]  
    \centering
    \begin{subtable}[c]{\textwidth}
        \caption{ }
        \centering
        \begin{tabular}{|c|c|c|c|c|c|c|c|c|c|c|}
            \hline Model & $s=2$ & 3 & 4 & 5 & 6  & 7 & 8 & 9 & 10 & 11    \\                
            \hline Empirical   & .880    & .749          & .639       & .527     & .431     & .355  & .273 & .218 & .159 & .110 \\
                                      & (.012)  & (.021)    & (.027)       & (.030)  & (.032)   & (.031) & (.030) & (.027) & (.024) & (.020) \\
            \hline Markov 1  &  .876     &  .761       & .659      &  .547     &  .445    & .365  & .283 & .219 & .163  & .116   \\
                                      & (.013)  & (.021)    & (.026)       & (.030)  & (.031)   & (.031) & (.030) & (.027) & (.024) & (.020) \\
            \hline Markov 2  &  .877     & .761        &  .658    &  .546     & .444      & .365  & .283 & .219 & .164 & .116  \\
                                      & (.013)  & (.021)    & (.026)       & (.030)  & (.032)   & (.031) & (.030) & (.027) & (.024) & (.020) \\
            \hline Markov 3  &  .876     &  .760       & .657     &  .545      &  .443    & .364  & .283 & .220 & .163  & .116   \\
                                      & (.013)  & (.021)    & (.026)       & (.030)  & (.032)   & (.031) & (.030) & (.027) & (.024) & (.020) \\
            \hline WT           &  .623     &  .406       & .269      &  .180   &  .122      & .083  &  .057 & .038 & .026 & .018   \\            
                                      & (.033)  & (.040)    & (.037)       & (.031)  & (.025)   &  (.020) &  (.015) & (.012) & (.009) & (.007) \\ 
            \hline
        \end{tabular}
        \vspace{5mm}           
    \end{subtable}
\quad%
    \begin{subtable}[c]{\textwidth}
        \centering
     \caption{}
        \begin{tabular}{|c|c|c|c|c|c|c|c|c|c|c|}
            \hline Model & $s=2$ & 3 & 4 & 5 & 6  & 7 & 8 & 9 & 10 & 11    \\                
            \hline Empirical   & .813    & .622         & .467       & .346      & .260     &.187    & .146  & .089 & .061 & .041  \\
                                      & (.023)  & (.035)    & (.041)       & (.043)  & (.042)   & (.039) & (.035) & (.031) & (.025) & (.021) \\
            \hline Markov 1  &  .810     & .644       & .512        & .395      & .292    & .216   & .158    & .113   & .077  & .052 \\
                                      & (.013)  & (.021)    & (.026)       & (.030)  & (.031)   & (.031) & (.029) & (.027) & (.024) & (.020) \\
            \hline Markov 2  &  .808     &  .642      &  .509       &  .392    & .291     & .216   & .158 &  .113 & .076  & .052 \\
                                      & (.013)  & (.021)    & (.026)       & (.030)  & (.032)   & (.031) & (.030) & (.027) & (.024) & (.020) \\
            \hline Markov 3  &  .808     &  .642      & .511        &  .393      &  .290  & 0.215 & .158 & .112 & .077  & .052  \\
                                      & (.013)  & (.021)      & (.026)     & (.030)  & (.032)   & (.031) & (.030) & (.027) & (.024) & (.020) \\
            \hline WT           &  .582     &  .356      & .224        &  .143   &  .092     & .059    &  .039   &.025    & .016   & .011   \\            
                                      & (.031)  & (.037)    & (.034)       & (.028)  & (.023)   &  (.018) &  (.014) & (.011) & (.008) & (.006) \\ 
            \hline
        \end{tabular}
        \vspace{5mm}           
    \end{subtable}
\quad%
\caption{Empirical and model based estimates of the probability of at
  least $s$ exceedances of the (a) 0.9 empirical quantile; (b) 0.95
  quantile; over a three week period given an exceedance of that
  threshold at the start of the period.\ Markov1-Markov3 are the
  correlation function models of the specified order and WT is the
  Winter--Tawn first-order model.\ Estimated standard errors are in
  parentheses.}
\label{AtLeastTab}
\end{table}

\begin{table} [htpb!]
    \centering
    \begin{subtable}[c]{\textwidth}
        \caption{ }
        \centering
        \begin{tabular}{|c|c|c|c|c|c|c|c|c|c|c|}
            \hline Model & $d=1$ & 2 & 3 & 4 & 5 & 6 & 7 & 8 & 9 & 10    \\                
            \hline Empirical   & .542    & .357       &  .285        & .265      & .235     & .222  & .190 & .187 & .188  & .181 \\ 
                                      & (.021)  & (.025)    & (.023)       & (.022)   & (.021)   & (.019) & (.019) & (.018) & (.017) & (.018) \\
            \hline Markov 1  & .533     &  .396       & .305       &  .302       &  .292    & .254    &  .206 & .201 & .216  &  .210  \\
                                      & (.025)  & (.031)    & (.031)       & (.030)   & (.026)   & (.023) & (.023) & (.020) & (.019) & (.020) \\
            \hline Markov 2  &  .590     & .410       &  .318      &  .298     & .269     & .278  & .216 & .220  & .265 & .228 \\
                                       & (.024)  & (.031)    & (.032)       & (.031)  & (.027)   & (.023) & (.023) & (.020) & (.019) & (.020) \\ 
            \hline Markov 3  &  0.557    & .399       & .298      &  .263      & .254   & .233  & .218  & .210  & .206  & .212   \\
                                      & (.025)  & (.032)    & (.032)       & (.031)  & (.026)   & (.023) & (.023) & (.020) & (.019) & (.020) \\
            \hline WT           &  .544    &  .361      & .252         &  .181   & .132     & .099 &  .073  & .054   & .041 & .031   \\            
                                      & (.024)  & (.028)    & (.027)       & (.025)  & (.023)   &  (.020) &  (.017) & (.015) & (.013) & (.011) \\ 
            \hline
        \end{tabular}
        \vspace{5mm}           
    \end{subtable}
\quad%
    \begin{subtable}[c]{\textwidth}
        \centering
     \caption{}
        \begin{tabular}{|c|c|c|c|c|c|c|c|c|c|c|}
            \hline Model &  $d=1$ & 2 & 3 & 4 & 5 & 6 & 7 & 8 & 9 & 10      \\                
            \hline Empirical   & .517    & .316       & .233         & .199      & .175     & .135  & .126  & .131 & .121  & .117  \\
                                      & (.030)  & (.036)    & (.037)      & (.036)  & (.036)   & (.030) & (.027) & (.024) & (.021) & (.019) \\
            \hline Markov 1  &  .509    & .362       & .264        & .257     & .228      & .186   & .155  & .160   & .142  & .114\\
                                      & (.033)   & (.038)    & (.036)       & (.034)  & (.030)   & (.025) & (.022) & (.020) & (.018) & (.016) \\
            \hline Markov 2  &  .513     & .364      &  .258        &  .236    & .205      & .206  &  .170 &  .178  &  .165   & .120   \\
                                      & (.035)  & (.042)     & (.040)      & (.037)  & (.032)   & (.026) & (.022) & (.020) & (.018) & (.016) \\
            \hline Markov 3  &  .517    &  .319      & .251        &  .240      & .188    &  .195    & .164 & .174   & .165  & .114  \\
                                      & (.034)  & (.044)     & (.040)     & (.037)    & (.031)   & (.026) & (.022) & (.020) & (.018) & (.016) \\
            \hline WT           &  .501    &  .315      & .212        &  .147     &  .105     & .075    & .055   &  .040  & .030  & .022  \\            
                                      & (.032)  & (.034)    & (.031)      & (.026)   & (.022)   &  (.018) &  (.014) & (.011) & (.009) & (.008) \\ 
            \hline
        \end{tabular}
        \vspace{5mm}           
    \end{subtable}
\quad%
\caption{Empirical and model based estimates of $\chi(v, d)$, defined
  in (\ref{SAE}), for $d$ from 1 to 10.\ In (a) the threshold $v$ is
  set to the 0.9 empirical quantile, while in (b) the 0.95 quantile
  was used.\ Markov1-Markov3 are the correlation function models of the
  specified order and WT is the Winter--Tawn first-order model.\
  Estimated standard errors are in parentheses.}
\label{ChiTab}
\end{table}

\subsection{Results}  \label{ResultsSec}

Throughout this section, $T$ will denote the function that transforms
daily maximum temperature in Orleans from the original scale
(\textdegree{}C) to the Laplace scale, as described in Section
\ref{MargMod}.\ Thus if $Y_t$ denotes the temperature in
\textdegree{}C in Orleans on day $t$ and $X_t = T(Y_t)$ then $X_t$ has
a standard Laplace distribution.

To get an initial impression for how daily maximum temperature in
Orleans evolves after exceeding 35\textdegree{}C, we may compute some
simple summary statistics via Monte Carlo simulation.\ For example,
suppose we want to estimate the expected maximum temperature over a
three week period conditional on 35\textdegree{}C being exceeded at
the start of the period.\ To do this we simulated $5\times\,10^5$
realizations, $\{\bm{X}^j_{1\,:\,21}\}_{j=1}^{5\times\,10^5}$ , of
$\bm{X}_{1\,:\,21}$ conditional on $X_1 > T(35)$ from our fitted
conditional extremes model by performing steps 1 to 6 of Algorithm
\ref{SAEI_Alg} with $d=21$, $n=5\times\,10^5$ and $v=T(35)$.\ We then
back transform to obtain $5\times\,10^5$ realizations,
$\{\bm{Y}^j_{1\,:\,21}\}_{j=1}^{5\times\,10^5} =
\{T^{-1}(\bm{X}^j_{1\,:\,21})\}_{j=1}^{5\times\,10^5}$, of
$\bm{Y}_{1\,:\,21}$ conditional on $Y_1 > 35$ and estimate
$\mathbb{E}(\text{max\,}\bm{Y}_{1:21}~|~Y_1 > 35)$ as
$(5\times\,10^5)^{-1}\sum_{j=1}^{5\times\,10^5}\text{max\,}\bm{Y}^j_{1:21}$.\
For the correlation function models of order 1, 2 and 3 we obtain
estimates of 37.00\textdegree{}C (0.36\textdegree{}C ),
36.98\textdegree{}C (0.38\textdegree{}C ) and 36.98\textdegree{}C
(0.37\textdegree{}C ) respectively, with estimated standard errors in
parentheses.\ For the Winter--Tawn model we obtain the estimate of
37.03\textdegree{}C (0.37\textdegree{}C ).\ Other summary statistics
may be computed similarly.

As was mentioned in Section \ref{MethModSec}, a period of three
consecutive days where the mean daily maximum temperature exceeds
35\textdegree{}C may lead to excess mortality in Orleans between 17\%
and 47\%.\ We estimated the probability of this event occurring during
a three week period conditional on 35\textdegree{}C being exceeded at
the start of the period from
\begin{equation}
  (5\times\,10^5)^{-1}\sum_{j=1}^{5\times\,10^5}\bigg(\mathbbm{1}\bigg[\sum_{i=1}^{19}\mathbbm{1}\bigg\{\frac{1}{3}\sum_{m=0}^{2}Y^j_{i+m} > 35 \bigg\} \geq 1 \bigg]\bigg).
\end{equation}
For the correlation function models of order 1, 2 and 3 this
probability is estimated as 0.457 (0.095), 0.445 (0.102) and 0.437
(0.105) respectively.  For the Winter--Tawn model we obtain an
estimate of 0.374 (0.076).

It is important that in the examples mentioned so far that estimates
of the quantities of interest are computed on the original scale,
i.e., using $\{\bm{Y}^j_{1\,:\,21}\}_{j=1}^{5\times\,10^5}$ rather
than $\{\bm{X}^j_{1\,:\,21}\}_{j=1}^{5\times\,10^5}$.\ We could not
compute an estimate on the Laplace scale and simply back transform.\
However, for some quantities of interest, estimates may be computed on
the Laplace scale without the need for back transforming.\ For
example, suppose we are interested in the probability that a
temperature of $s$\textdegree{}C is exceeded over a three week period
given that 35\textdegree{}C is exceeded at the beginning of the
period.\ Then, since
$\mathbb{P}(\text{max\,}\bm{Y}_{1:21} > s~|~Y_1 >35) =
\mathbb{P}(\text{max\,}\bm{X}_{1:21} > T(s)~|~X_1 >T(35))$, we may
estimate this quantity directly using Algorithm \ref{SAEI_Alg} with
$v=T(35), d=21$ and
$g(\bm{x}) = \mathbbm{1}\big[\max\bm{x}_{1\,:\,21} > T(s) \big]$.\ We
used Algorithm \ref{SAEI_Alg} with $v=T(35), d=21$ and
$n=5\times\,10^5$, for the following choices of the function $g$,
\begin{align}
  g_s(\bm{x}) &=  \mathbbm{1}\big[\max\bm{x}_{1\,:\,21} > T(s) \big],  \label{MaxExc}  \\
  g_s(\bm{x}) &=  \mathbbm{1}\bigg[ \sum_{i=1}^{21}\mathbbm{1}[x_i > T(35)] \geq s \bigg],  \label{TotExc}   \\
  g_s(\bm{x}) &=  \mathbbm{1}\bigg[\max\, \bigg\{{i\in\,1{\,:\,}20}\Bigm| \min\bm{x}_{t+1\,:t+i} > T(35)\,\,\text{for some\,}  t\in\,0{\,:\,}(21-i)\bigg\} \geq s \bigg].    \label{ConsecExc}
\end{align}
All of the functions $g$ in (\ref{MaxExc})-(\ref{ConsecExc}) depend on
a single parameter $s$.\ Taking $s=36$ for example in (\ref{MaxExc})
would yield via Algorithm \ref{SAEI_Alg} an estimate of the
probability that 36\textdegree{}C is exceeded during a three week
period given an exceedance of 35\textdegree{}C at the beginning of the
period.\ The function $g$ in (\ref{TotExc}) is used to estimate the
probability of at least $s$ exceedances over the three week period,
whereas (\ref{ConsecExc}) is used to estimate the probability of at
least $s$ consecutive exceedances.\ Estimates of
$\mathbb{E}\{g_s(\bm{x})~|~X_1 > T(35)\}$ for each of
(\ref{MaxExc})-(\ref{ConsecExc}) are shown in Table
\ref{ClusterFuncsEst} for various choices of the parameter $s$.\ The
correlation function models of all orders provide similar estimates
for each function $g$ and choice of the parameter $s$.\ For estimating
the probability of exceeding the temperature $s\textdegree{}$, i.e.,
with $g_s$ as in (\ref{MaxExc}), the Winter--Tawn model produces
similar estimates as the correlation models but smaller estimates for
the other functions.\ However, most of the Winter--Tawn estimates are
within one or two standard errors of our correlation model estimates.\
Thus although the diagnostics from Section \ref{DiagSec2} suggest that
our models perform better at lower thresholds, there is little in the
way of a significant difference at the much higher threshold
considered in this section, consistent with the simulation results
reported at the end of Section \ref{ExSemiParam}.

In applications, as is done in \cite{winttawn16}, it may be desirable
to report estimates for expectations of functionals such as
(\ref{MaxExc})-(\ref{ConsecExc}) over clusters of extremes as opposed
to over a fixed block length as we do here.\ As pointed out in Section
\ref{ForSim}, this requires only a trivial modification to the method
of this section.\ To achieve this latter modification, we simply
truncate each of our simulated blocks
$\{\bm{X}^j_{1\,:\,21}\}_{j=1}^{5\times\,10^5}$ so that it corresponds
to a cluster, e.g.\,using the runs method \citep{smithweiss94}, and
then compute any quantities of interest using these clusters.\ The
block length of 21 is sufficiently long that the probability of a
cluster exceeding this value is negligible, however if this were not
the case then a larger block length may be used.
\begin{table}[htbp!]
\centering
\caption{Model based estimates for
  $\mathbb{E}\{g_s(\bm{X}_{1\,:\,21})~|~X_1>T(35)\}$ obtained via
  Algorithm \ref{SAEI_Alg}.\ Markov1-Markov3 are the correlation
  function models of the specified order and WT is the Winter--Tawn
  first-order model.\ Estimated standard errors are in parentheses.}
 \label{T1}
\begin{tabular}{|c | c | c | c | c | c |}
  \hline 
  Function  &  $s$   &  Markov 1 &  Markov 2 & Markov 3 & WT  \\
  \hline
            &        $36$      &  $.6998\, (.0747)$      &   $.6966\,(.0763)$   &  .6970\,(.0757) & .6804 \,(.0627)\\
            &        $37$      &  $.4350\, (.0968) $     &   $.4297\,(.0998)$   &  .4270\,(.0992) & .4295\,(.0877) \\
  $g_s(\bm{x})$ in (\ref{MaxExc})                        &        $38$      &  $.2259\, (.0796) $       &   $.2211\,(.0832)$    &   .2206\,(.0825) & .2483\,(.0831)  \\   
            &        $39$      &  $.1043\, (.0538) $     &   $.1007\,(.0571)$   &  .1004\,(.0566) & .1272\,(.0622) \\   
            &        $40$      &  $.0464\, (.0323)$      &   $.0443\,(.0348)$   &  .0445\,(.0344) & .0542\,(.0409) \\   
            &        $41$      &  $.0147\, (.0181)$      &   $.0142\,(.0197)$   &  .0142\,(.0194) & .0182\,(.0243) \\
            &        $42$      &  $.0014\, (.0076)$      &   $.0013\,(.0086)$   &  .0013\,(.0085) & .0040\,(.0133) \\
  \hline
            &          $ 2 $   &  $.6269\,(.0949)$       &   $.6108\,(.0984)$   &  .6118\,(.0959) & .5031\,(.0632) \\
            &          $3$     &  $.4040\,(.0908)$       &   $.3967\,(.0975)$   &  .3904\,(.0985) & .2716\,(.0638) \\
            &          $4$     &  $.2583\,(.0716)$       &   $.2525\,(.0767)$   &  .2493\,(.0764) & .1513\,(.0517) \\   
            &          $ 5 $   &  $.1650\,(.0526)$       &   $.1626\,(.0562)$   &  .1605\,(.0560) & .0859\,(.0389) \\   
  $g_s(\bm{x})$ in (\ref{TotExc})                   &         $ 6  $   &    $.1119\,(.0379)$      &   $.1103\,(.0396)$  &  .1093\,(.0396)  & .0494\,(.0284) \\   
            &         $ 7  $   &  $.0712\,(.0265)$       &   $.0708\,(.0271)$   &  .0705\,(.0272) & .0286\,(.0205) \\  
            &         $ 8  $   &  $.0482\,(.0191)$       &   $.0481\,(.0190)$   &  .0487\,(.0190) & .0164\,(.0147) \\   
            &         $ 9  $   &  $.0330\,(.0137)$       &   $.0331\,(.0136)$   &  .0334\,(.0135) & .0097\,(.0105) \\  
            &         $ 10 $  &   $.0197\,(.0094)$       &   $.0192\,(.0092)$   &  .0199\,(.0092) & .0058\,(.0075) \\  
            &         $ 11 $  &   $.0120\,(.0068)$       &   $.0120\,(.0066)$   &  .0122\,(.0066) & .0033\,(.0054) \\
  \hline
            &           $2$     & $.5310\,(.0923)$       &   $.5115\,(.0978)$   & .5163\,(.0935)  & .4629\,(.0657) \\
            &           $3$     & $.2834\,(.0755)$       &   $.2742\,(.0840)$   & .2642\,(.0872)  & .2297\,(.0612) \\
            &           $4$     & $.1513\,(.0503) $      &   $.1470\,(.0570)$   & .1417\,(.0574)  & .1187\,(.0464) \\   
            &           $ 5 $   & $.0864\,(.0320)$       &   $.0843\,(.0365)$   & .0823\,(.0371)  & .0631\,(.0331) \\   
  $g_s(\bm{x})$ in (\ref{ConsecExc})                   &           $ 6 $   &    $.0466\,(.0217)$      &   $.0457\,(.0241)$     & .0453\,(.0241) & .0340\,(.0231)  \\   
            &         $ 7  $   &  $.0319\,(.0161)$       &   $.0314\,(.0169)$   & .0320\,(.0170)  & .0185\,(.0160) \\  
            &         $ 8  $   &  $.0222\,(.0122)$      &    $.0220\,(.0122)$   & .0225\,(.0122)  & .0101\,(.0111) \\  
            &         $ 9  $   &  $.0151\,(.0092)$      &    $.0151\,(.0092)$   & .0155\,(.0092)  & .0057\,(.0078) \\  
            &         $ 10 $   &  $.0101\,(.0066)$       &   $.0100\,(.0065)$   & .0105\,(.0064)  & .0032\,(.0055) \\  
            &          $11$    &  $.0066\,(.0049)$      &    $.0067\,(.0047)$   & .0068\,(.0047)  & .0017\,(.0039) \\
  \hline
\end{tabular}   \label{ClusterFuncsEst}
\end{table}

\clearpage

\appendix

\section{Proof of Theorem 3.1.}  \label{AppendixProof}
\begin{proof}
To show unbiasedness, 
\begin{align}
\mathbb{E}_{\pi^{*}}\bigg\{\frac{g(\bm{X}_{1\,:\,d})}{S(\bm{X}_{1\,:\,d})}\bigg\} & = \int \frac{g(\bm{x})}{S(\bm{x})}\pi^{*}(\bm{x})d\bm{x}
 = \frac{1}{\bar{p}}\int_{\mathcal{L}} \frac{g(\bm{x})}{S(\bm{x})}\bigg(\sum_{j=1}^{d} \mathbbm{1}_{L_j}(\bm{x})\pi(\bm{x}) \bigg)d\bm{x} \notag \\
& = \frac{1}{\bar{p}}\int_{\mathcal{L}} \frac{g(\bm{x})}{S(\bm{x})}S(\bm{x})\pi(\bm{x})d\bm{x}  \notag \\
& = \frac{1}{\bar{p}}\int_{\mathbb{R}^d} g(\bm{x})\pi(\bm{x})d\bm{x}  
 =  \frac{1}{\bar{p}}\mathbb{E}_{\pi}\{g(\bm{X}_{1\,:\,d})\},  \label{ExpForm_Cond}
\end{align}
where (\ref{ExpForm_Cond}) follows as $g$ is supported on
$\mathcal{L}$.\ Linearity of expectation then shows that
$\widehat{\mathbb{E}}_{\pi}\{g(\bm{X})\}$ is unbiased.

By a similar calculation we find that
\begin{equation}
\mathbb{E}_{\pi^{*}}\bigg\{\frac{g(\bm{X}_{1\,:\,d})^2}{S(\bm{X}_{1\,:\,d})^2}\bigg\} = \frac{1}{\bar{p}}\int_{\mathcal{L}} \frac{g(\bm{x})^2}{S(\bm{x})}\pi(\bm{x})d\bm{x}
\end{equation}
and so 
\begin{equation}
  \text{var}\bigg\{\frac{g(\bm{X}_{1\,:\,d})}{S(\bm{X}_{1\,:\,d})}\bigg\} = \frac{1}{\bar{p}}\int_{\mathcal{L}} \frac{g(\bm{x})^2}{S(\bm{x})}\pi(\bm{x})d\bm{x} - \bigg(\frac{\mathbb{E}_{\pi}\{g(\bm{X}_{1\,:\,d})\}}{\bar{p}}\bigg)^2.
\end{equation}
Using independence and (\ref{owens3}) gives the formula for the
variance as claimed.

To prove the upper bound (\ref{UpperBound_Cond}), we note that if $g$ is an indicator function, then $g(\bm{x})^2 = g(\bm{x})$ and so 
\begin{equation}
\int_{\mathcal{L}} \frac{g(\bm{x})^2}{S(\bm{x})}\pi(\bm{x})d\bm{x} = \int_{\mathcal{L}} \frac{g(\bm{x})}{S(\bm{x})}\pi(\bm{x})d\bm{x} \leq \int_{\mathcal{L}} g(\bm{x})\pi(\bm{x})d\bm{x} 
=  \int_{\mathbb{R}^d}g(\bm{x})\pi(\bm{x})d\bm{x}   = \mathbb{E}_{\pi}\{g(\bm{X})\}
\end{equation}
as $S \geq 1$ on $\mathcal{L}$ and using this bound gives the result.

Finally, to see that $\widehat{\mathbb{E}}_{\pi}\{g(\bm{X})\}$ is a
consistent estimator when $g$ is bounded, we just need to check that
\begin{equation}
\left\lvert \int_{\mathcal{L}} \frac{g(\bm{x})^2\pi(\bm{x})}{S(\bm{x})}d\bm{x} \right\rvert < \infty
\end{equation}
which follows easily from the fact, if
$\lvert g(\bm{x}) \rvert < C < \infty$ and $S(\bm{x}) \geq 1$, then
$\int_{\mathcal{L}} g(\bm{x})^2\pi(\bm{x})/S(\bm{x})d\bm{x} \\< C^2$
as $\pi$ is a density function.
\end{proof}

\clearpage

\section{Diagnostic plots}  \label{AppendixDiag}

\begin{figure}[h]
  \centering
  \includegraphics[scale=0.55]{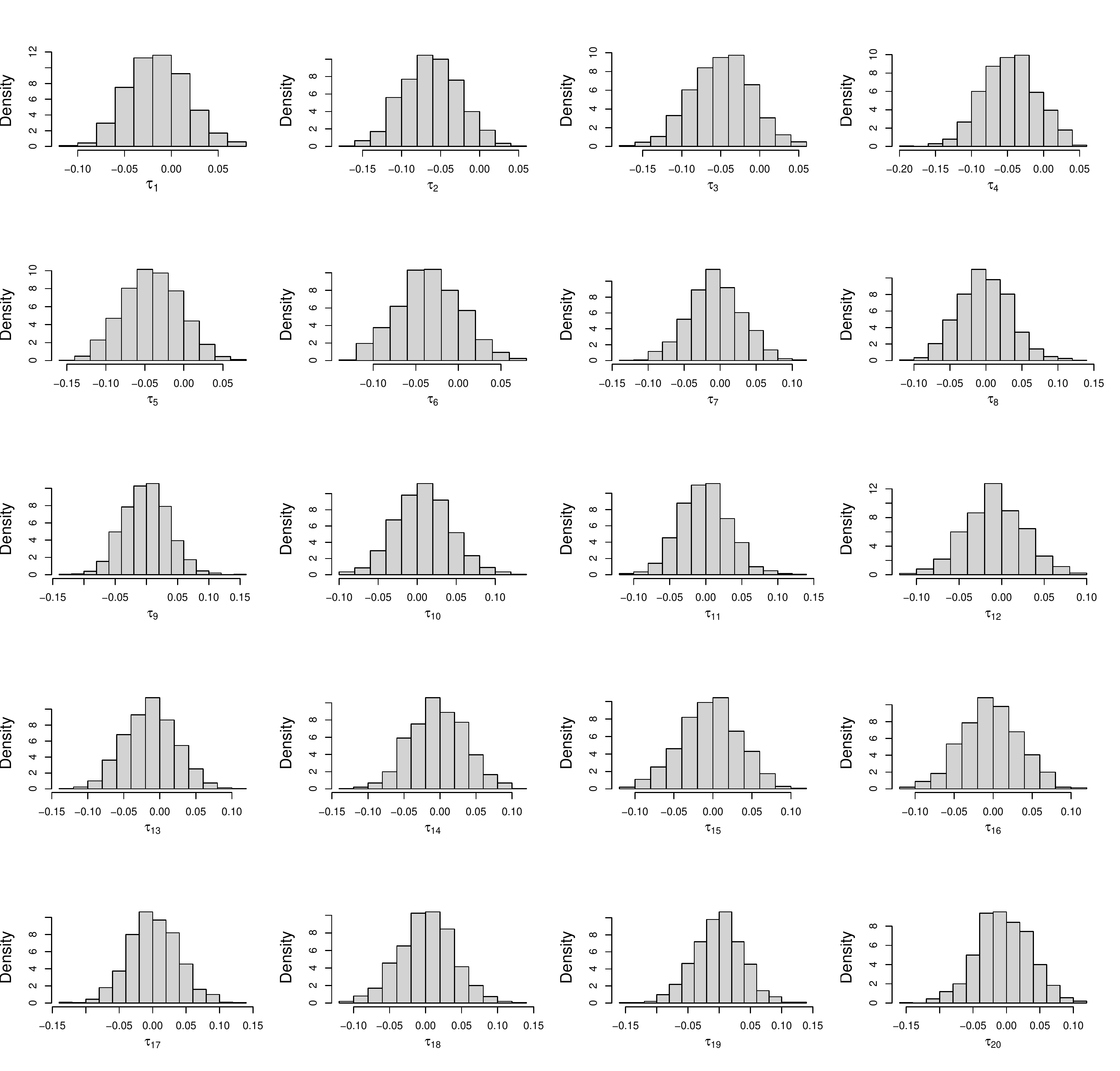}  
  \caption{Bootstrap distributions of $\tau_i$, $i\in\,1{\,:\,}20,$
    for the first-order correlation function model, where $\tau_i$ is
    Kendall's measure of association between
    $\{(X_t, \hat{Z}_{t+i\mid t})\}_{t\in\,T_u}$.}
\label{KendallTauBoot}
\end{figure} 

\clearpage

\textbf{Acknowledgements:} This paper is based on Chapter 3 of
\cite{AuldThesis}.\ The authors thank Jonathan Tawn and Miguel de
Carvalho for helpful comments that helped improve this paper.

{\small
\bibliographystyle{agsm}
\bibliography{TSCE_AuldPapas.bib}
}

\end{document}